\documentclass[prd,showpacs,superscriptaddress]{revtex4}
\usepackage{epsfig}

\begin{document}

 \title{The  Concepts of ``Age''  and ``Universality''
 in Cosmic  Ray Showers}

\author{Paolo Lipari}
\email{paolo.lipari@roma1.infn.it}
\affiliation{INFN  sez. Roma ``La Sapienza''}

\begin{abstract}
The concept of ``age'' as a parameter 
for the description of the    state of  development
of  high energy showers in the atmosphere
has been in use  in cosmic  ray  studies for  several decades. 
In  this  work we briefly  discuss     how this concept,
originally  introduced to  describe the average
behavior of  electromagnetic  cascades,  can be  
fruitfully applied
to  describe  individual  showers   generated by  primary
particles  of different  nature, including protons, nuclei and
neutrinos.
Showers with  the same age share three different
important properties: (i)  their electron size  has the same 
fractional rate of   change  with increasing depth,
 (ii) the  bulk of  the electrons and  photons in  the shower 
(excluding  high  energy particles)  have energy spectra
with shapes  and  relative normalization uniquely
determined by  the age parameter, 
(iii) the electrons and photons
in the shower  have also  the same angular  and
 lateral distributions  sufficiently far from the shower axis.
In this  work we  discuss how  the properties
associated with the shower  age can be  understood 
with simple arguments, and how the shapes
of the  electron and photon spectra  and the  relative
normalization   that correspond to a   certain age
can be calculated  analytically.
 \end{abstract}



\maketitle

\section{Introduction}
The concept of the ``age'' of a shower  has  been
in use in the cosmic  ray community 
for more  than half a  century. The concept  first 
emerged \cite{Rossi-Greisen}  in the study   of the average 
longitudinal   development 
of purely  electromagnetic  showers  generated   by photons
or electrons. It was then also  applied  
\cite{Nishimura-Kamata,Greisen1} to
the lateral distribution of electrons around the shower axis.
Soon,  it was also understood that  it is possible and 
useful to assign an ``age'' also to individual  showers,
and that the concept is  applicable also to showers
generated  by hadronic   primary particles  such as protons
or nuclei.

Some  recents  works \cite{Giller,Nerling,Gora}
have   rediscussed the  concept of of shower  age 
for  the showers  generated  by ultra high energy 
cosmic rays in the Earth's atmosphere. 
Giller et al   \cite{Giller}  and Nerling et al. \cite{Nerling} 
have studied with  montecarlo methods the showers generated  by
high energy  protons  and  nuclei  in air,
and  have  observed  that the energy and angle 
 distributions  of the electrons   (in this  work with
``electrons''  we   will refer to  the sum of electrons and positrons)
 in the showers
 have shapes that to a   good  approximation  are only determined  by 
 an  age  parameter  $\overline{s}$    defined  as:  
 \begin{equation}
 \overline{s} (t, t_{\rm max} ) 
= \frac{3 \,t}{t + 2 \; t_{\rm max}} ~.
 \label{eq:s0}
 \end{equation}
where  $t$ is  the depth  in unit of radiation lengths
and  $t_{\rm max}$ is the 
depth where the shower  reaches its maximum  size.
The shower size  is defined  as
the  total number of charged particle 
integrated over all energy, and effectively coincides with the
electron size.
These results have been extended to the lateral distribution
of  electrons by Gora et al. \cite{Gora}.
This property of ``universality''  is obviously very  important for the 
analysis and interpretation 
of  high energy cosmic  ray  observations,
and it is therefore  desirable to have a deeper understanding 
of its origin and   of its limitations.

In this   work we want to review  critically the  concepts of
shower ``age''  and ``universality''.  One of the main
points we want to   make 
is to   argue that the  definition
of age of equation (\ref{eq:s0}), while 
reasonably accurate in most cases, 
is in general not correct, and should be  replaced  
by a    better  motivated and more accurate definition.

The  essence of the   concept of shower age  can be 
understood observing that    all  showers  at the 
maximum of their development are  in  an appropriate sense
``similar''  to each other  (that is have the same ``age'').
This ``similarity'' is  represented by the  fact that in all
showers at maximum   the energy spectra of 
 ``most''  electrons and  photons
have   the  same  shape and  the same relative normalization.
These particles  also have  the same  angular  distributions
(that are  obviously strongly correlated with energy)
and, for  an equal  density profile of the medium
where the  shower is propagating, also the same lateral
distribution around the shower  axis.

The   idea  that all  showers at  maximum,  
that is at the  depth where the   derivative  of the
shower size $N(t)$ vanishes, are ``similar''
independently from the   energy
and nature of the primary particle, 
can be naturally be generalized, stating that 
all  showers   that have the same
fractional    rate of  change with depth, 
that is the same ``size slope'' $\lambda$:
\begin{equation}
\lambda =  \frac{1}{N(t)} \; \frac{dN(t)}{dt}
\end{equation}
are also ``similar''. This  means
that to each  value of the  $\lambda$ correspond
well determined shapes of the electron and  photon spectra
(again only valid   for ``most''  particles) and  a   well determined
relative normalization for the two populations.
It is  intuitive
(and will be later  verified by  detailed calculations) 
that    the  energy  spectra of electrons and
photons  become  progressively  softer 
as   $\lambda$ decreases     going from   positive 
(when the shower  size  grows)  to  negative 
(when the shower size decreases)   values.

There is a  one to one mapping 
between the   values of the size slope  $\lambda$  and    the
values of the shower ``age'' $s$.  This  mapping
is   encoded  by  a  function that, in the notation
introduced by Rossi and Greisen  in their ``classic'' 
paper \cite{Rossi-Greisen}, is  called  $\lambda_1(s)$.
The     general  definition of the shower age is therefore:
\begin{equation}
s = \lambda_1^{-1} ( \lambda)  = \lambda_1^{-1} \left (
\frac{1}{N(t)} \; \frac{dN(t)}{dt}
\right )
\label{eq:age-def}
\end{equation}
where  $\lambda_1^{-1}$ is the inverse function  of $\lambda_1(s)$.
The   function $\lambda_1(s)$ (that  will be discussed
in more detail in the following) is  monotonically
decreasing and   has a single  zero   at $s = 1$,
therefore   according
to equation (\ref{eq:age-def})  
showers  have   age   $s=1$ at maximum  and  age 
$s < 1$ ($s >1$)   before  (after) maximum.

The definition of  the age parameter (\ref{eq:age-def})   may seem 
at first sight (and in some  sense actually is)
arbitrary,   since the   size slope  $\lambda$ itself
or any  monotonic  function  of $\lambda$ 
are also   perfectly  adequate to  identify ``similar''  showers.
The choice of the particular mapping 
of    equation (\ref{eq:age-def})
is   motivated by the fact that  one can attach
a  direct physical  meaning   to
the  quantity $s$. The shapes  of the electron
and photon spectra   for  $E$ above the critical energy
$\varepsilon$
(the electron  critical energy 
$\varepsilon$ is the average
energy lost by an electron in a radiation length, and corresponds
also to the electron energy for which radiative and collision 
losses are equal)
 and  $E \ll E_0$    (with $E_0$ the primary
particle energy)  are  well  represented by a power law:
\begin{equation}
n_e (E) \sim n_\gamma (E) \sim E^{-(s+1)}
\end{equation}
These power  law  behaviors  stops when 
$E$ approaches (from above) the electron critical 
energy. 
For  energies  below $E \sim \varepsilon$
the electron spectrum has a sharp cutoff, 
while the photon spectrum  
has a ``knee''  and  takes the form $E^{-1}$.  
The precise  shapes of the  cutoff of the electron spectrum,
of  the knee of the  photon spectrum 
(that is the transition    from
the  form $E^{-(s+1)}$ to the   $E^{-1}$)
and  the relative normalizations of the photon and electron 
spectra  are  all entirely determined  by the age  $s$ (or 
equivalently  by the size slope $\lambda$)
and can  be computed in   detail.

The  commonly used   definitions  of  age  
in equation (\ref{eq:s0}) 
always coincides  with the general  definition (\ref{eq:age-def})
at shower maximum 
and therefore,   by construction, it is a   reasonably good
approximation  for  showers   sufficiently
close to maximum.  
The motivation for the more  general   definition may  
appear as only a   formal question of  ``principle''.
In fact in some  circumstances the  two definitions
are significantly different, and the general  definition
gives the correct  shower age. 
The definition (\ref{eq:s0})  coincides  with the  correct one 
only for a particular  shape of  the shower
longitudinal  that  is known as   the 
``Greisen  profile'' \cite{Greisen1}.
In fact  Greisen profile
and the age definition (\ref{eq:s0}) 
are intimately connected, 
and can be seen  (with the mediation  of the function $\lambda_1$) 
as  the integral  and the  derivative of each other.
The Greisen profile (discussed   below  in section~\ref{sec:greisen})
describes accurately the average development of   
purely electromagnetic  showers,
but  is only a    rough approximation  for  the  description 
of individual  hadronic showers, and naturally fails  completely
in the description of neutrino--induced showers.
The deviations of the  definition (\ref{eq:s0}) 
from the true age  (\ref{eq:age-def})
are of  the same order of the deviations of
the profile of a shower from  the Greisen  profile
that has the same $t_{\rm max}$.

The authors  in \cite{Giller,Nerling}    
have  calculated with montecarlo methods 
the shape of the electron spectra  
in hadronic showers  of different age.
The parametrizations of their results  are   essentially
identical  to the  shapes  of  the electron 
spectra   in  showers of the same  age
calculated    several  decades  ago  by
Rossi and Greisen \cite{Rossi-Greisen}. 
These  modern  works  have therefore 
effectively only ``rediscovered'' with 
montecarlo methods    what should  be called
the ``Rossi--Greisen''  spectra. 
We want to  attract attention to this   fact  for  three reasons.
The first one is  that it  is  obviously appropriate   to give
credit  to  the  remarkable work of  the  pioneers.
The second is that 
the works of   \cite{Giller,Nerling} 
do  not include a discussion of   photon spectra.
The shapes of these spectra
and their relative normalization  with respect to the  electron ones
are also  determined  unambiguously by the shower  age, and have
 been also  computed explicitely  by Rossi and Greisen.
Finally the   derivation 
of   the spectral shapes  obtained by Rossi and Greisen
with analytic  methods allows   physical  insights
on  the  origin  and  limitations  of the ``universality'' 
of the spectra,  that are not    easily deducible from 
a  montecarlo calculation.

It should be stressed   that  the ``universality''
of properties for  cascades  of the same age has  clearly 
 limitations  since it only applies
to  ``most''  but not all  particles
in  the  shower.  
For example,
the ``similarity''  among showers 
at   the maximum of their  development,
(that is at age $s=1$)
does not imply that the showers 
simply differ   by the  absolute normalization  of 
their  electromagnetic  component. 
Considering at first the case of purely electromagnetic cascades,
at maximum the  showers  generated
by a photon of initial  energy $E_0$
contain (in essentially all cases)
 more high energy particles   than  
the showers generated by photons of lower energy.
These   high energy particles are negligible  
in number   and  do not 
contribute   significantly  to the total size
but in general  carry an important fraction of the shower energy,
they  ``feed''  the shower development 
and  influence its  development. 
The showers generated by other  types
of  primary particles   
have ``cores''  of  different  structure
and particle content,  and    follow different   development profiles.

\vspace{0.3 cm}
This  work is  organized as follows:  
in the next   two sections  we review  a
very well known  subject, discussing   the average 
longitudinal  evolution of  purely electromagnetic  showers
first in ``approximation~A'',  that is  neglecting
the  electron ionization losses, and then in ``approximation~B''.
The concept of age  emerged   naturally   in these studies. 
In approximation~B   the shower equations have ``elementary solutions''
labeled  by the   parameter $s$, these  solutions 
 correspond to the ``universal spectra'' of showers
with age $s$.
The following  section  discusses 
the well known ``Greisen profile''
that  describes the average longitudinal development
of  purely  electromagnetic showers.   
Finally we   discuss the evolution of individual   
hadronic  showers, and  give some  conclusions.

\section{Electromagnetic  Showers  in Approximation~A}
The   evolution of  purely electromagnetic  showers
can be studied 
\cite{Rossi-Greisen}
using two  sets of  simplifying assumptions
called ``Approximation A''  and ``Approximation B''.

In approximation~A  the  only  processes   
considered  for the    shower  development
 are  pair production 
  for  photons,  and   bremsstrahlung   for   electrons.
The  differential cross sections
for  these processes  are  described by
the  asymptotic  formulae valid at high energy.  
The electron energy losses   due to collisions  with 
electrons  and nuclei of the medium  are neglected.

The  average longitudinal  
development of  electromagnetic  showers
is described
by  the two functions $n_e(E,t)$ and $n_\gamma(E,t)$ 
that  give the  differential energy spectra
of electrons and photons  
at  depth $t$.
In  this  work we are  following   the  notation
introduced by Rossi and Greisen in \cite{Rossi-Greisen},
however  here    we  introduce  different     symbols.
Rossi and Greisen indicate 
the  differential  (integral) electron  spectrum  
as $\pi (E,t)$ ($\Pi(E,t)$)
and the photon spectrum as  $\gamma(E,t)$;
the subscript  notation used here is more suitable to 
the extension  of the formalism
to  hadronic  showers  where other 
particle types are present.

In approximation~A the  evolution of the 
electron  and photon  differential spectra 
is    described 
 by the two   integro--differential equations:
 \begin{eqnarray}
 \frac{\partial n_e (E,t)}{\partial t} & =  &
 -  \int_0^1 dv~ \; \varphi_0(v) \;
 \left [n_e (E,t) - \frac{1}{1-v} \, n_e 
  \left (  \frac{E}{1-v}, t \right)
 \right ] 
 + 2 \; \int_0^1 \frac{du}{u} \; \psi(u) \;
 n_\gamma \left (  \frac{E}{u}, t \right) ~,
 \label{eq:show1}
 \\
 & ~ &\nonumber \\
 & ~ &\nonumber \\
 \frac{\partial n_\gamma (E,t)}{\partial t} 
 & = &
 \int_0^1 \frac{dv}{v} \; \varphi(v) \;
 n_e \left (  \frac{E}{v}, t \right)
 - \sigma_0 \; n_\gamma (E,t) 
 \label{eq:show2} ~.
 \end{eqnarray}
 In   the right hand side of
 equation (\ref{eq:show1})   the first  term
 describes the $(e\to  e)$  contribution, and the 
second one  the   ($\gamma \to e$)  processes.
 In   the right hand side of
  equation (\ref{eq:show2})   the first   term
 describes the $(e\to  \gamma)$   contribution,  and
 the second one photon absorption.
 The   differential cross sections for 
 bremsstrahlung $\varphi(v)$ and   
pair production $\psi(u)$, and the
 photon absorption cross section $\sigma_0$ are 
given in appendix A.

 \subsection{Elementary solutions}
\label{sec:sola}
\begin{figure} [hbt]
\centerline{\psfig{figure=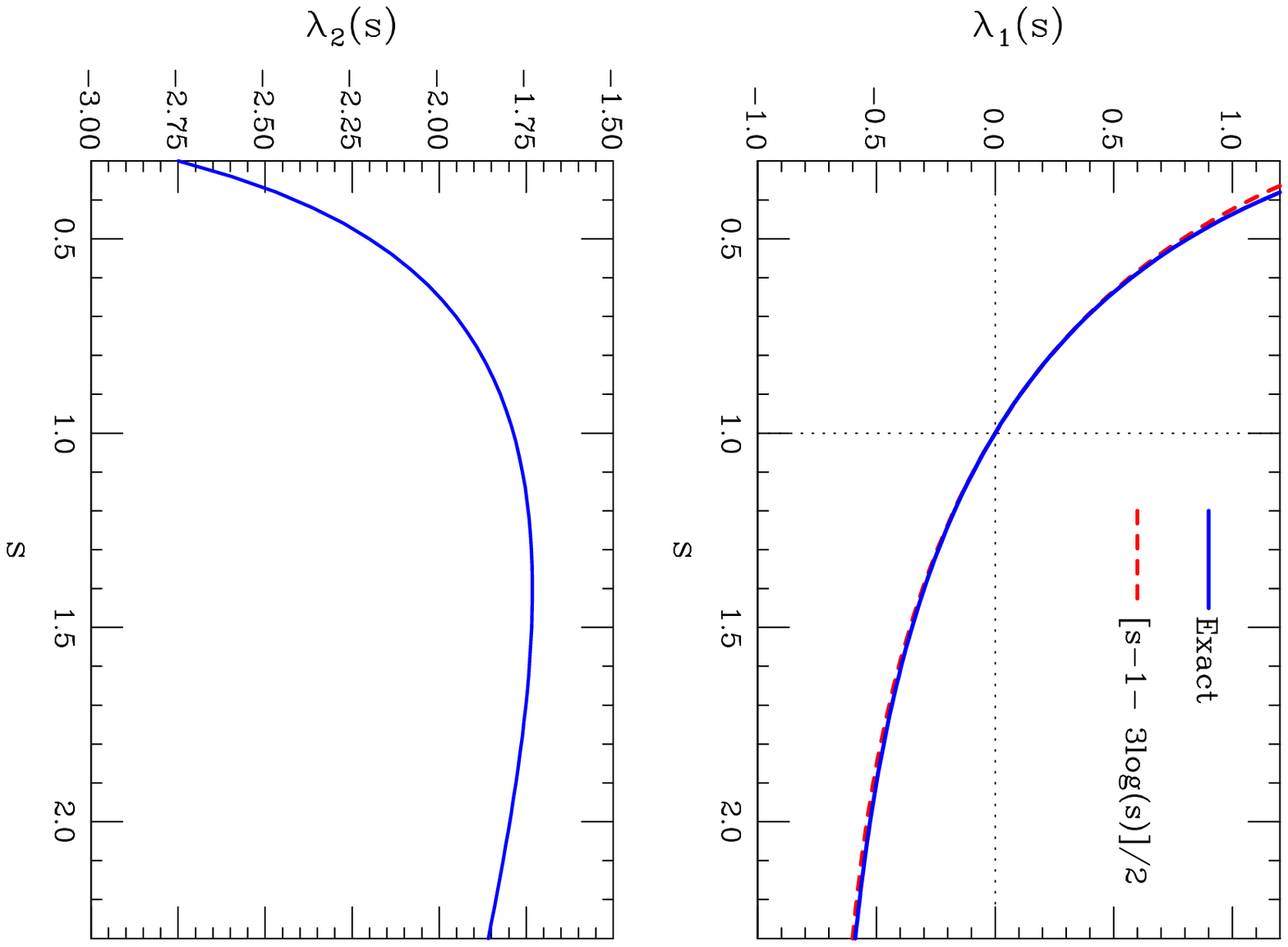,angle=90,width=16.0cm}}
\caption {\footnotesize 
Top panel: plot of the function $\lambda_1(s)$; the dashed line
shows the   analytic approximation introduced by Greisen.
Bottom panel:
plot of the function $\lambda_2(s)$.
\label{fig:lambda}  }
\end{figure}

\begin{figure} [hbt]
\centerline{\psfig{figure=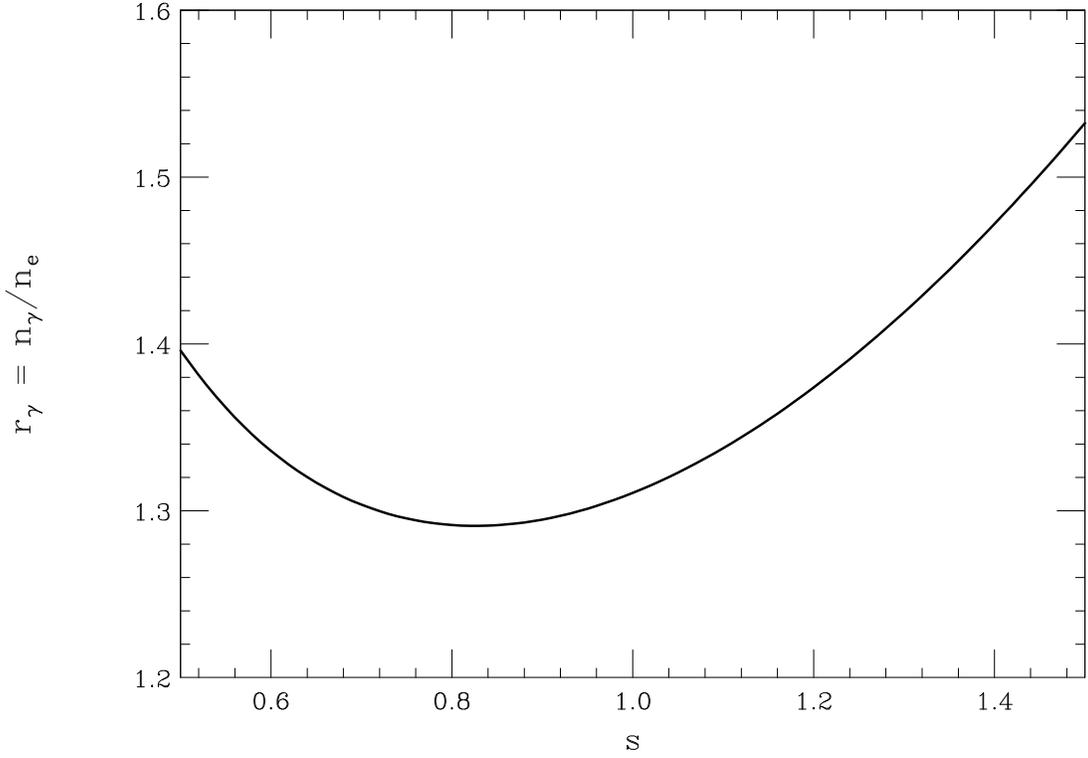,angle=90,width=14.5cm}}
\caption {\footnotesize 
Equilibrium photon/electron ratio:
$r_{\gamma}^{(1)} (s) = C(s)/[\sigma_0 + \lambda_1(s)]$.
\label{fig:rge}  }
\end{figure}

 In the system of equations 
 (\ref{eq:show1}) and (\ref{eq:show2})     no energy scale is  present.
Accordingly   these equations have a set 
 of  ``elementary'', scale  invariant  solutions of  form:
 \begin{equation}
 \left \{ 
 \begin{array}{l c l} 
 n_e (E, t) & = &  K \; E^{-(s+1)} ~e^{\lambda(s) \, t} \\[0.3 mm]
 n_\gamma (E, t) & = & 
  K \; r_{\gamma} (s) \; E^{-(s+1)} ~e^{\lambda(s) \, t} 
 \end{array}
 \right .
 \label{eq:pow0}
 \end{equation}
 that  are   power laws   in  energy, and 
 change  exponentially  with  the depth $t$.
 Inserting these  solutions  in the
 shower equations (\ref{eq:show1}) and (\ref{eq:show2}) one  
 obtains   a  quadratic  equation for 
 $\lambda(s)$ that has the  two solutions:
 \begin{equation}
 \lambda_{1,2} (s) =
 - \frac{1}{2} \left (A(s) + \sigma_0 \right )
 \pm  \frac{1}{2}\sqrt{ \left (A(s) - \sigma_0 \right )^2 +
 4 \, B(s) \, C(s)}.
 \label{eq:lambda}
 \end{equation}
To  each  solution corresponds a 
photon/electron ratio:
 \begin{equation}
 r_{\gamma}^{(1,2)}(s) = \frac{C(s)}{\sigma_0 +  \lambda_{1,2}(s)}
 \label{eq:rgamma}
 \end{equation}
 The  auxiliary  functions   $A(s)$, $B(s)$ and $C(s)$  
 appearing in  the definitions
 (\ref{eq:lambda})  and  (\ref{eq:rgamma}) 
are  given  explicitely  in appendix A.
The  functions   $\lambda_{1,2}(s)$ 
are shown in  fig.~\ref{fig:lambda},
the function $r_{\gamma}^{(1)}$ is shown in fig.~\ref{fig:rge}.

Even if  $\lambda_1(s)$ is already given by  an explicit 
analytic expression in (\ref{eq:lambda}),
it is  very useful to  
follow  Greisen \cite{Greisen1}  and 
introduce the  simpler  expression
 \begin{equation}
\overline{\lambda}_1 (s) =
  \frac{1}{2} \left (
 s -1 - 3 \, \ln s \right )
 \label{eq:lambda1-greisen}
 \end{equation}
 that  is  a  very good approximation 
for $\lambda_1(s)$  with deviations smaller
 than 2\% in the interval $0.6 \le s \le 1.4$. 
 A  comparison of the exact and approximate expressions
 for $\lambda_1(s)$ is shown in the top panel
 of  fig.~\ref{fig:lambda}.
The usefulness of this  simpler  functional  form will
be  clear in the following.

The existence of  two solutions  $\lambda_{1,2}(s)$ 
for each  $s$  value  is   physically  simple to  understand.
If one starts at $t=0$    with  populations of 
electrons   that have  power law  form    with the
same slope, but   arbitrary  normalizations,
  the  spectra  
mantain   identical power law shapes at all  $t$,  
but change their relative and absolute normalization.
The spectra  reach first an asymptotic    $\gamma/e$ ratio
with a $t$ scale  $|\lambda_2(s)|^{-1}$, and then 
evolve   exponentially   $\propto e^{\lambda_1(s) \, t}$
mantaining a  constant  equilibrium ratio.
The  convergence to an asymptotic  $\gamma/e$ ratio
is the   fastest  process  since 
 $|\lambda_2 (s)| > |\lambda_1 (s)|$  for
 all $s$ values.

 As an explicit example,  
an  initial  power law,
pure  electron spectrum:
 \begin{equation}
\left \{
\begin{array}{l l l}
 n_e(E,0) & = & K ~E^{-(s+1)}
 \\[3 mm]
 n_\gamma(E,0) & = &  0
\end{array}
\right .
 \end{equation}
evolves in $t$ as:
 \begin{equation}
\left \{
\begin{array}{l l l}
 n_e(E,t) & = & \frac{K}{\lambda_1(s) - \lambda_2(s)} 
 ~\left [ 
 (\lambda_1(s) + \sigma_0)  \; e^{\lambda_1(s) \, t} 
 -(\lambda_2(s) + \sigma_0)  \; e^{\lambda_2(s) \, t} 
 \right ]
 ~E^{-(s+1)}
  \\[3 mm]
 n_\gamma(E,t) & = & \frac{K}{\lambda_1(s) - \lambda_2(s)} 
 ~C(s) ~\left [ 
 e^{\lambda_1(s) \, t} 
 -e^{\lambda_2(s) \, t} 
 \right ]
 ~E^{-(s+1)}
\end{array}
\right .
 ~.
 \label{eq:example-power}
 \end{equation}
The spectra  remain power  laws for all $t$  values.
For $t \gg |\lambda_2(s)|^{-1}$    one can 
set $e^{\lambda_2(s) \, t}$ to zero, and the spectra
evolve in $t$ as a simple  exponential  
($\propto   e^{\lambda_1(s) \, t}$) with 
an  asymptotic $\gamma/e$ ratio 
$C(s)/(\lambda_1(s) + \sigma_0)$
that corresponds to the first solution in (\ref{eq:rgamma}).

In summary,  the  solutions 
 \begin{equation}
\left \{
\begin{array}{l l l}
 n_e(E,t) & = & K^\prime ~E^{-(s+1)} ~e^{\lambda_1 (s) \,t}
 \\[3 mm]
 n_\gamma(E,0) & = & K^\prime ~E^{-(s+1)} ~e^{\lambda_1 (s) \,t}~r_\gamma^{(1)}(s)
\end{array}
\right .
\end{equation}
is a sort of   ``attractor'', and  any combinations
of photon and electron spectra  power laws  spectra
of the same  slope  $s$  ``converge''  to this solution.

The  only $t$ independent 
solution  corresponds to  $s=1$  and  is  particularly important:
The existence  of  this $t$--independent    solution:
 \begin{equation}
 \left \{ 
 \begin{array}{l c l} 
 n_e (E, t) & = &  K \; E^{-2}  \\[0.3 mm]
 n_\gamma (E, t) & = &  K \; \overline{r}_{\gamma} \; E^{-2} 
 \end{array}
 \right .
 \label{eq:scale}
 \end{equation}
and its  energy  dependence $\propto E^{-2}$ 
do not  depend on the detailed form of 
 the pair production and bremsstrahlung cross sections,
and  can be immediately understood 
observing  that a power  law   spectrum of   form
 $E^{-2}$   contains   equal  amount of  energy   in each
 energy decade and  that the 
 bremsstrahlung  and pair production processes
 that ``mix'' the electron  and  photon  populations, 
  conserve  energy and are scale  invariant.
  Only the value
 of the  $\gamma/e$ ratio   that corresponds
to this solution
 depends on the detailed  form  of the 
 cross sections and is:
 \begin{equation}
 \overline{r}_{\gamma} = 
 r_{\gamma}^{(1)} (1)  
  =  \frac{\langle v\rangle_{\rm brems}}{\sigma_0}
=  \frac{C(1)}{\sigma_0}
 = \frac{(1+b)}{(7/9 -  b/3)}  \simeq 1.31
 \end{equation}

The  fact  that $\lambda_1(s)$ is  positive 
(growing solution)  for  $s <1$
and   negative  (decreasing solution)  for  $s >1$
 is also independent  from the  detailed form
 of the cross sections,
 and is  a simple consequence
 of the fact that  in a power  law  spectrum of form
 $E^{-(s+1)}$   the energy  contained  in each decade
 increases  with $E$  when    $s < 1$  
 and  decreases when  $s> 1$.

This   very  elementary  discussion  already illustrates
how the $t$--dependence of  the shower development
is  intimately related to the  shape of  the  energy spectra of the 
particles in the shower.

\subsection{Showers generated by a primary   $\gamma$ or $e^\mp$.}
The evolution of the shower   generated by a primary
electron or photon of  initial  energy $E_0$,
in approximation~A has  
been  calculated   by  Rossi and Greisen \cite{Rossi-Greisen}.
Before  describing these  solutions  explicitely
we can  observe that some  important properties
can be  readily deduced with 
simple  considerations.  
The   differential spectra of electrons and photons
in the solutions have  the   scaling form:
\begin{equation}
n_\alpha (E_0, E, t) =
 \frac{1}{E_0} \; f_\alpha \left ( \frac{E}{E_0}, t \right ) 
\end{equation}
(the  subscript $\alpha$  runs over the 4 cases:
$e\to e$, 
$e \to \gamma$,
$\gamma \to e$ and 
$\gamma \to \gamma$).
Correspondingly   the integral spectra have 
the scaling form:
\begin{equation}
N_\alpha (E_0, E_{\rm min}, t) =
\int_{E_{\rm min}}^{E_0} dE~n_\alpha(E_0,E, t)= 
 F_\alpha \left ( \frac{E_{\rm min}}{E_0}, t \right )~.
\end{equation}
These  scaling properties of  the approximation~A solutions
are  a simple 
consequence of the absence  of
quantities with the dimension of energy in the shower equations.

Energy conservation  is reflected in the
condition:
\begin{equation}
\int_0^{E_0} dE ~E ~ n_e (E_0, E,t) + 
\int_0^{E_0} dE ~E ~ n_\gamma (E_0, E, t)  = E_0 
\label{eq:en-cons-A}
\end{equation}

The solution of the shower  equations for a monochromatic
photon or electron  \cite{Rossi-Greisen}   is  very simple
for the  the Mellin  transforms  of   $n_e$ and $n_\gamma$.
The physically observable 
spectra can then  be obtained inverting
these transforms.  The inversion   can be  easily 
performed numerically with   a   single path   integration 
along a line  in  the complex plane, with exact results.
Rossi  and Greisen have  also
shown that it is possible to  invert  the trasform  
using a  ``saddle point approximation'', obtaining
simple  expressions for the spectra  that are at the same  time
remarkably accurate and very instructive.

\begin{figure} [hbt]
\centerline{\psfig{figure=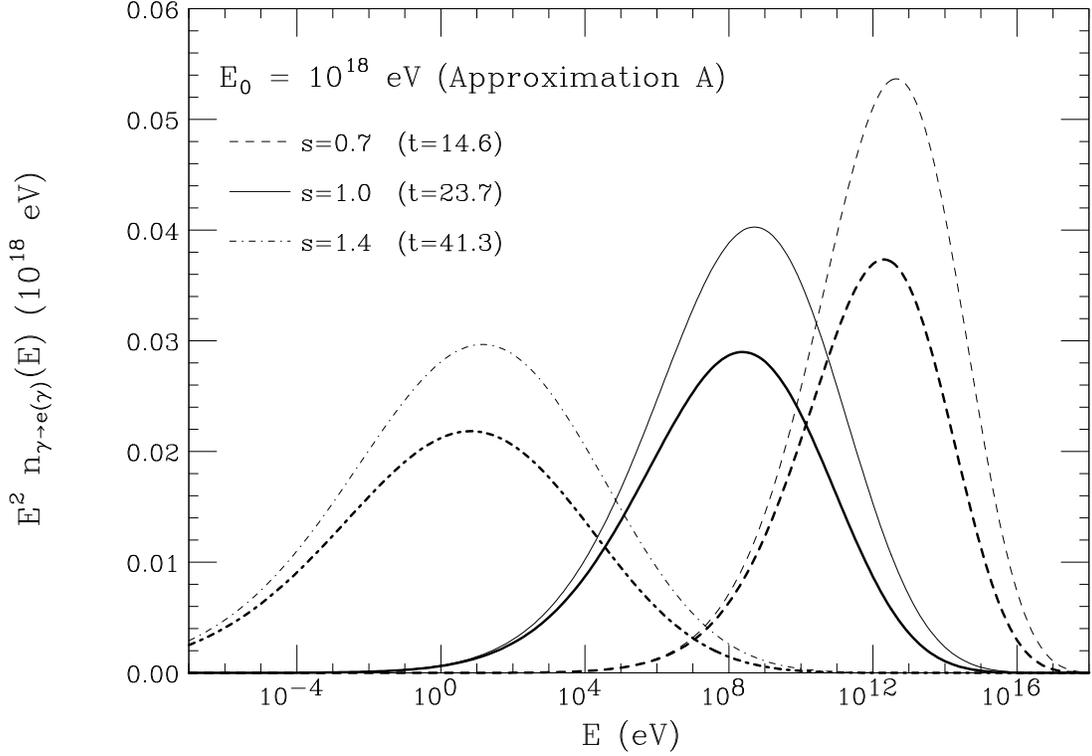,angle=90,width=14.5cm}}
\caption {\footnotesize 
Electron and  photon  energy spectra  
calculated in  approximation A for the  shower  generated 
by a primary photon of  energy $10^{18}$~eV
at three  values of the depth ($t=14.6$, 23.7,  t=41.3
that correspond approximately  to  age $s=0.7$, 1 and 1.4).
Thick (thin) lines are for electrons (photons). 
The spectra are  shown in the form 
$E^2 \; n(E)$  versus $E$.  
The area  below each curve 
is  proportional to the amount of  energy   transported by
each particle type at the depth  considered.
\label{fig:appra}  }
\end{figure}

The  saddle point approximation  solution
for the differential spectra
(valid  for  large    $t$ and  $E/E_0 \ll 1$)
can be written as:
\begin{equation}
n_{\alpha} (E_0, E, t)  \simeq
\frac{1}{E_0} ~ \frac{1}{\sqrt{2 \pi}}~\left [ 
\frac {G_\alpha (s)}{\sqrt{ \lambda_1^{\prime \prime}(s) \, t}}
~ \left ( \frac{E}{E_0}  \right )^{-(s+1)} ~
e^{\lambda_1 (s) \, t}
\right ]_{s = \overline{s} (E/E_0, t) }
\label{eq:approxa}
\end{equation}
with
\begin{equation}
\overline{s} \left ( \frac{E}{E_0}, t \right ) \simeq
\frac{3t}{t - 2 \ln(E/E_0)} ~.
\label{eq:sa}
\end{equation}
For  the integral  spectra the saddle point approximation
solution is:
\begin{equation}
N_{\alpha} (E_0,E_{\rm min}, t) 
   \simeq 
\frac{1}{\sqrt{2 \pi}}~\left [ \frac{1}{s} \;
\frac {G_{\alpha} (s)}{\sqrt{ \lambda_1^{\prime \prime}(s) \, t}}
~ \left ( \frac{E_{\rm min}}{E_0}  \right )^{-s} ~
e^{\lambda_1 (s) \, t}
\right ]_{s = \overline{s}(E_{\rm min}/E_0, t) }
\label{eq:nint0}
\end{equation}
Note that both the  electron and photon  integral  spectra   
diverge  for  $E_{\rm min} \to 0$.
For  completeness 
a derivation  of these  well known    results
is  sketched in appendix B,
that also  lists  explicit  expressions for the   functions 
$G_{\alpha}(s) $.
An example of the    differential spectra 
in approximation A is shown in  fig.~\ref{fig:appra}.

The   saddle point  solutions of the shower equations
in approximation~A   exhibit  several interesting properties: 
\begin{itemize}

\item
For a   fixed value (less than  unity) of the ratio  $E/E_0$
(or $E_{\rm min}/E_0$) the differential and integral spectra  
start at zero  for  $t \simeq 0$, then   grow  with increasing $t$,
reaching  a maximum  at  the value $t_{\rm max}$
and then  begin to decrease  vanishing for $t\to \infty$.
The exponential   factor $e^{\lambda_1 (s) \, t}$ 
controls  the $t$ evolution of the solution.
Therefore  in good approximation  shower maximum 
corresponds  to $\lambda_1(s) = 0$ (and therefore to  $s=1$).
Using equation (\ref{eq:sa}) one  finds the  well known  result:
\begin{equation}
t_{\rm max} \left ( \frac{E}{E_0}\right )  \simeq
\ln \left ( \frac{E_0}{E}\right )
\end{equation}
On can therefore  rewrite equation (\ref{eq:sa}) in the form:
\begin{equation}
\overline{s} \left ( \frac{E}{E_0}, t \right ) \simeq
\frac{3t}{t + 2 \, t_{\rm max}} ~.
\label{eq:sb}
\end{equation}

\item 
More in general,  from the fact that the factor
$e^{\lambda_1 (s) \, t}$  controls the $t$ dependence  of  the
solution, it   follows  that to a good approximation one has:
\begin{equation}
\frac{1}{N_\alpha}
 \; \frac{\partial N_{\alpha}}{\partial t} 
\simeq \lambda_1 \left [\overline {s} 
\left ( \frac{E_{\rm min}}{E_0}, t \right ) \right ]
\label{eq:longa1}
\end{equation}
or
\begin{equation}
\frac{1}{n_\alpha}
 \; \frac{\partial n_{\alpha}}{\partial t} 
\simeq \lambda_1 \left [\overline {s} 
\left ( \frac{E}{E_0}, t \right ) \right ]
\label{eq:longa2}
\end{equation}

\item The quantity $\overline{s} (E/E_0, t)$
is   related to the  shape of the 
energy spectrum  around  $E$, and has  manifestly 
 the meaning of  the ``local slope''  
of the spectrum   at energy $E$:
\begin{equation}
-\frac{E}{n_{\alpha}} \;
\frac{\partial n_{\alpha}}{\partial E} \simeq 
\overline{s} 
\left ( \frac{E}{E_0}, t \right ) + 1
\end{equation}
In other words the spectrum around $E$ is  well approximated
 as  a power law $E^{-(\overline{s}+1)}$.
Globally the spectrum is  not  a simple power law
and  the ``local slope''  changes as a function of  $E/E_0$ and $t$.
For  a fixed depth $t$
the local slope $\overline{s}$  grows  monotonically 
with $E/E_0$, as the  spectrum
becomes   progressively steeper.
For a fixed  value of  $E/E_0$  the  spectrum
also  becomes  steeper with increasing $t$  and
as the local slope   grows  monotonically.
The shape  of the spectrum around  energy $E$ 
  takes  the form $\propto E^{-2}$ 
when  the spectrum for this energy 
reaches the maximum.

\item  Photons and electrons  have spectra
with  very similar  but  not identical  shapes,
and accordingly the $\gamma/e$ ratio    changes  slowly  with
energy. From  the expressions for the 
functions $G_\alpha(s)$ 
(given in (\ref{eq:g1}--\ref{eq:g4}))
one finds:
\begin{equation}
\frac{n_\gamma (E_0,E,t)}{n_e(E_0,E,t)} 
\simeq  \frac{C(\overline{s})}{\sigma_0 + \lambda_1 (\overline{s})}
= r_\gamma^{(1)} (\overline{s})
\label{eq:rra}
\end{equation}
That is  the ``asymptotic ratio''  for the elementary
power law  solution with  slope $s = \overline{s}(E/E_0,t)$.
Note the remarkable fact that this result is independent 
from   the nature of the   particle  ($\gamma$ or $e$)
that  initiates the shower.

\item In fact the   developments of   showers  initiated
by a photon or an electron of the same  energy 
are  remarkably  close to each  other, 
demonstrating  that the  electron and photon population
quickly tend to reach a  sort of dynamic  equilibrium 
feeding each other
\end{itemize}

In approximation~A  there is  no  natural way
to associate a  single age to a shower  because the
equations do  not  contain 
any meaningful energy  scale
(except the energy of the primary particle). 
Therefore  for a given     depth 
$t$  one can associate a different 
age    to   every    ratio  $E/E_0$
(or $E_{\rm min}/E_0$)  according to
equation  (\ref{eq:sa}).
 The  age $\overline{s}$
describes at the same  time
the ``stage''  of the
longitudinal evolution of  spectrum
(via  equations (\ref{eq:longa1}) 
or
(\ref{eq:longa2})),
and  the shape  of the spectrum
near energy $E$, that is  well approximated by
the  power  law $\propto  E^{-(\overline{s} +1)}$.
The age  also controls the $\gamma/e$ ratio
around $E$ according to  equation (\ref{eq:rra}).

\section{Electromagnetic  Showers  in Approximation~B}
In approximation B the electron  energy losses 
due to  collisions are simply  modeled 
 as  an energy independent  loss
 $\varepsilon$ per   unit of radiation length.
The quantity $\varepsilon$ is the critical  energy
(in air  $\varepsilon \simeq 81$~MeV).
Accordingly, 
 a   term is added  to the right hand side of  
equation  (\ref{eq:show1})  that describes the electron
evolution:
 \begin{eqnarray}
 \frac{\partial n_e (E,t)}{\partial t} & =  &
 -  \int_0^1 dv~ \; \varphi_0(v) \;
 \left [n_e (E,t) - \frac{1}{1-v} \, n_e 
  \left (  \frac{E}{1-v}, t \right)
 \right ] 
 \nonumber \\
 & ~ & \label{eq:show1b} \\
 & ~ & 
 + 2 \; \int_0^1 \frac{du}{u} \; \psi(u) \;
 n_\gamma \left (  \frac{E}{u}, t \right)
+  \varepsilon \; \frac{\partial n_e (E,t)}{\partial E}
 \nonumber 
 \end{eqnarray}
The  new system of  equations 
[(\ref{eq:show1b})  and (\ref{eq:show2})]
does not 
have any more 
simple power  law  solutions  of form (\ref{eq:pow0}).
However, even in this case, 
one can  introduce  ``elementary''  solutions 
   that  have
 a constant shape  in energy and  evolve  with $t$ with the
 simple behavior  $e^{\lambda(s) \, t}$.
 Following Rossi and Greisen  \cite{Rossi-Greisen} 
  the  elementary solutions can be written  in the form:
 \begin{equation}
 \left \{ 
 \begin{array}{l c l} 
 n_e (E, t) & = &  
  K \;   \; e^{\lambda(s) \, t} 
 \; E^{-(s+1)}
 \; p\left (s, \frac{E}{\varepsilon} \right ) 
\\[0.3 mm]
 n_\gamma (E, t) & = & 
  K \;   e^{\lambda(s) \, t} 
\; E^{-(s+1)}
\; g\left (s, \frac{E}{\varepsilon} \right ) 
\; r_{\gamma} (s) ~ \; 
 \end{array}
 \right .
 \label{eq:pow1}
 \end{equation}
 that  contain   two additional  functions  $p(s,x)$ and
 $g(s,x)$.
 For large  energy ($E \gg  \varepsilon$) the 
  electron collision losses  can be  safely  neglected,
 and the solutions
 coincide with the  simple power law form
 of approximation A.
 This constraint tell us  that the functions
 $\lambda(s)$ and $r_{\gamma}(s)$
that  appear  in (\ref{eq:pow1}) coincide 
with the  functions discussed  before and  given in equations
 (\ref{eq:lambda})  and (\ref{eq:rgamma}),
and that   for large $E/\varepsilon$
the functions $p(s,x)$ and $g(s,x)$ 
asymptotically become unity:
 \begin{equation}
 \lim_{x\to\infty} p(s,x) = 1,
 ~~~~~~ 
 \lim_{x\to\infty} g(s,x) = 1.
 \end{equation}

 Inserting expression 
 (\ref{eq:pow1})  in the shower equations  
 one obtains  two  pairs of   integro-differential  equations
 (see appendix C)
 for the functions $p(s,x)$ and $g(s,x)$ 
  corresponding to the two solutions  for $\lambda(s)$.
 These equations can be solved  
 numerically to  obtain the functions $p_{1,2}(s,x)$ 
 and $g_{1,2}(s,x)$.

The physical  meaning of the functions 
$p_{1}(s,x)$   and $g_{1}(s,x)$  is   trasparent.
If one injects  power laws   spectra of  electrons  and
photons of form $E^{-(s+1)}$
after  a  few lengths $|\lambda_2(s)|^{-1}$
the   spectra  take  asymptotically  constant  shapes
given  by:
 \begin{equation}
 \left \{ 
 \begin{array}{l c l} 
 n_e (E, t) & = &  
  K \;   \; e^{\lambda_1(s) \, t} 
 \; E^{-(s+1)}
 \; p_1\left (s, \frac{E}{\varepsilon} \right ) 
\\[0.3 mm]
 n_\gamma (E, t) & = & 
  K \;   e^{\lambda_1(s) \, t} 
\; E^{-(s+1)}
\; g_1\left (s, \frac{E}{\varepsilon} \right ) 
\; r_{\gamma}^{(1)} (s) ~ \; 
 \end{array}
 \right .
 \label{eq:pow2}
 \end{equation}
(identical to  (\ref{eq:pow1}) but selecting the  first 
of the two possible solutions)
 and  continue  their evolve
in $t$  as a simple  exponential.

The qualitative features  of this asymptotic solution
are  easy to understand. The electron spectrum is
a nearly perfect  power  law  for $E \gg \varepsilon$  
but  has a  cutoff for  $E \sim \varepsilon$, 
when electrons are  absorbed  because of ionization losses.
The photon spectrum  changes   from a power law  
$\propto E^{-(s+1)}$
for  $E \gg \varepsilon$ to the form
$\propto E^{-1}$ for  $E \ll \varepsilon$, reflecting
the $1/E$   dependence of the bremsstrahlung cross section.

These physically intuitive properties are  confirmed by
the explicit calculation  first  performed by Rossi and Greisen,
who   have demonstrated  \cite{Rossi-Greisen}
that  the  behavior of the  functions 
$p(s,x)$ and $g(s,x)$ for 
 $x \to 0$ is:
 \begin{eqnarray}
 p(s,x) & \propto   &  x^{s+1}
  \label{eq:ser1}
  \\
 g(s,x) & \propto   &  x^s  
  \label{eq:ser2}
 \end{eqnarray}

 The  low  energy behavior of  the   function
 $p(s,x)$  implies that  
for   $E \to 0$  the   electron spectrum  goes to
a  finite value. 
The energy integration of the  electron spectrum 
therefore converges both
for   $E\to \infty$ (if  $s >0$)
and for $E\to 0$, and it becomes possible to    talk about
the total  electron size.
As  expected the  differential photon spectrum
diverges $\propto E^{-1}$ for   $E\to 0$, and therefore
the   integral spectrum  diverges  logarithmically
at the lower limit.

\begin{figure} [hbt]
\centerline{\psfig{figure=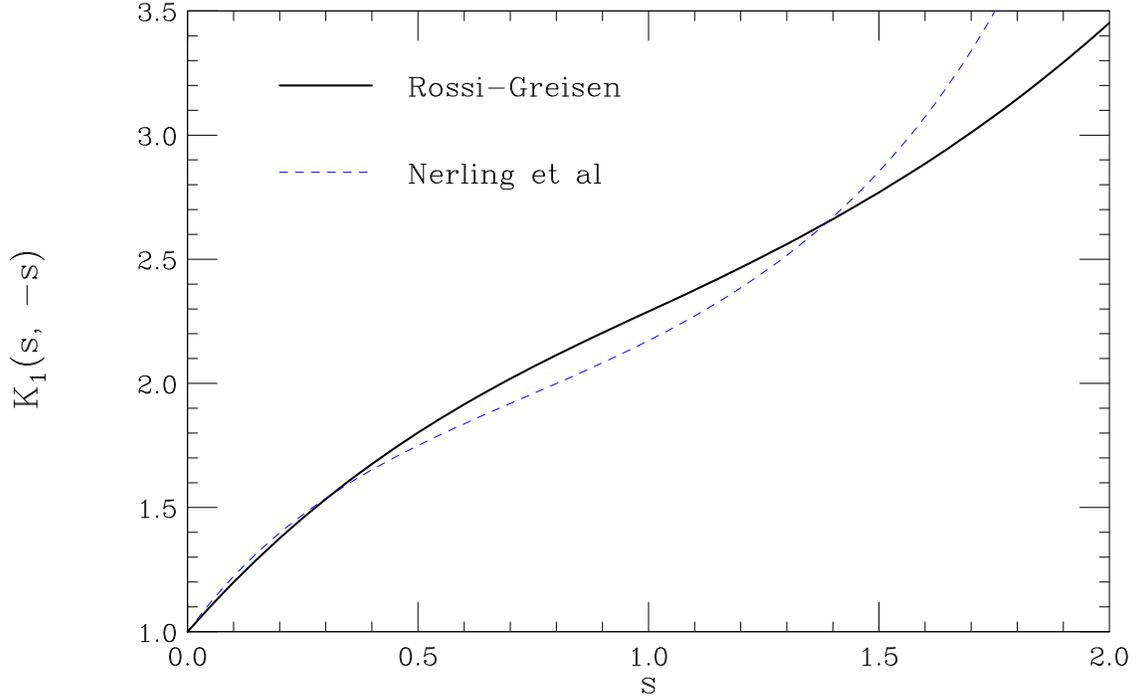,angle=90,width=15cm}}
\caption {\footnotesize 
Plot of the function $K_1(s,-s)$. 
The dashed line
shows the   normalization of the electron spectrum
of Nerling et al \cite{Nerling}.
\label{fig:ks}  }
\end{figure}

The total  electron size   for the
phenomenologically most  important solution (that
corresponds to $\lambda_1 (s)$) 
can  be written as:
\begin{eqnarray}
N_e (s)  =  \int_0^\infty dE ~n_e (E)  & \propto &
  \int_0^\infty dE  ~E^{-(s+1)} ~ p_1 \left ( s,\frac{E}{\varepsilon} 
\right )  \nonumber \\
& ~ & \nonumber  \\
& = & \varepsilon^{-s} ~ 
  \int_0^\infty dx  ~x^{-(s+1)} ~ p_1 (s,x)=
 \varepsilon^{-s} ~\frac{K_1 (s,-s)}{s}
 \label{eq:pint}
\end{eqnarray}
The last equation   defines   the function $K_1(s,-s)$.
 The  physical  significance  of  $K_1(s,-s)$ 
 can  be understood  comparing equation 
 (\ref{eq:pint}) with the integral 
 $$
 \int_{\varepsilon}^\infty  dE  ~E^{-(s+1)} = \frac{\varepsilon^{-s}}{s}
 $$
 The  electron size  obtained  integrating over  all
$E$   the   elementary solution (\ref{eq:pow1})  differs from the  
 integration of  the simple form $E^{-(s+1)}$ 
 in the interval $(\varepsilon \le E \le \infty$) by
 a  factor $K_1(s,-s)$.
Rossi and Greisen  have  shown how to calculate
exact values  of the  function $K_1(s, -s)$  
for all integer values  $s \ge 0$. For $s=1$ one has
$K_1(1,-1) = 2.8948$.
A   plot of the function $K_1(s,-s)$ is shown in fig.~\ref{fig:ks}.

The  functions  $p_{1,2}(s,x)$ and $g_{1,2}(s,x)$ 
 can be calculated  with numerical methods,
as discussed in Appendix C.
For $x > 1$, it is also possible  \cite{Rossi-Greisen}
 (see again appendix~\ref{sec:expansion})  to  express the functions
as power  expansion in  $1/x$  
with easily calculable  coefficients.

\begin{figure} [hbt]
\centerline{\psfig{figure=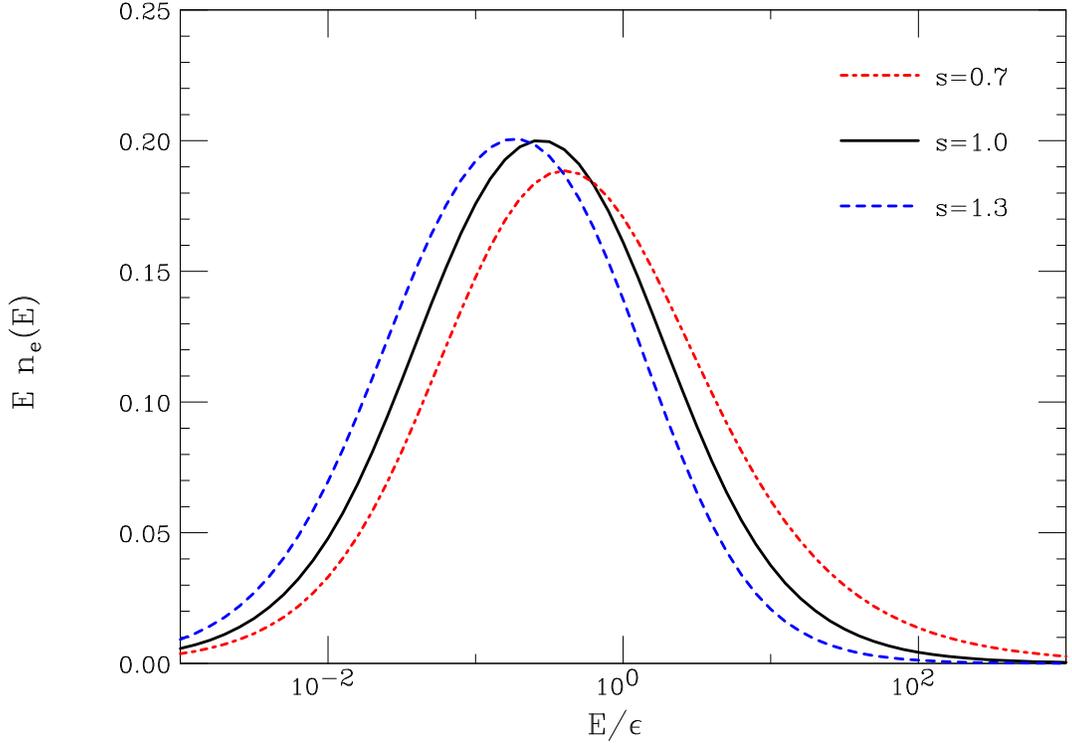,angle=90,width=14.2cm}}
\caption {\footnotesize 
Plot of the energy distributions of electrons 
calculated in approximation~B 
for three values of the age parameter $s$
($s=0.7$,
$s=1$  and 
$s=1.3$).
The  distributions 
(in the form   $E \, n_e = dn_{e}/d\ln E$) 
are calculated as:
$p_1(s,E) \; E^{-s}$ and 
are renormalized to have  a total  size of one electron.
\label{fig:sp1}  }
\end{figure}

\begin{figure} [hbt]
\centerline{\psfig{figure=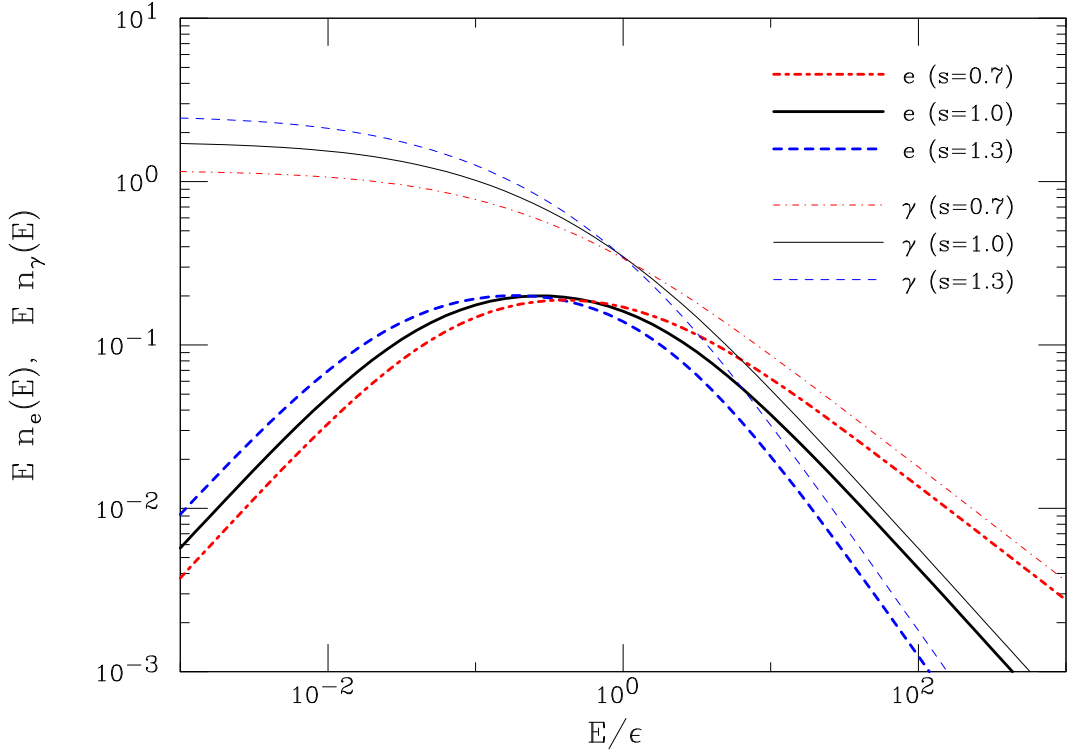,angle=90,width=14.2cm}}
\caption {\footnotesize 
Plot of the energy distributions of electrons  and photons
calculated in approximation~B 
for three values of the age parameter $s$
($s=0.7$, 1  and  1.3)
The distributions
(in the form  $E \, n_{e\gamma} = dn_{e,\gamma}/d\ln E$) 
are calculated as:
$p(s,E) \; E^{-s}$  for  electrons 
$g(s,E) \; E^{-s} \, r_{\gamma}(s)$ for  photons
and renormalized to have  a total  size of one electron
(and the  correct $\gamma/e$ ratio)
\label{fig:sp2}  }
\end{figure}

 Figures~\ref{fig:sp1}  and~\ref{fig:sp2}   show
  (on a linear and log   scale)
 the  behavior of  the  electron 
 spectrum   plotted in the form 
$dn_e/d\ln E = E \, n_e (E)$ for 
 three values of  the index  $s$. 
 The   integrated  electron size $N_e$
 is  accounted for  by particles in  
the energy range
 $0.01 \lesssim E/\varepsilon \lesssim 10$.
Note that  an important limitation  of the 
treatment in approximation B is  that  the
electron mass is  neglected. Accordingly  
the electron  spectra extends down to $E\to 0$ 
to   unphysical  energy values  below the electron mass.

\subsection{Solutions  for  monochromatic electron or photon}
The  solution   of the shower  equations
 in approximation B   with the  initial condition
of a monochromatic  electron or  photon of
energy $E_0$  cannot  be  given with an exact  closed form
expression. 
Rossi and  Greisen  suggest to
approximate   the solution with  the expressions:
\begin{eqnarray}
n_{e(\gamma) \to e} (E_0, E,t)  & \simeq   &
\left [ n_{e(\gamma) \to e} (E_0, E,t) \right ]_A \times
p_1 \left [\overline{ s} \left (\frac{\varepsilon}{E_0}, t \right ), 
 \frac{E}{\varepsilon}\right ]
\\
& ~ & \nonumber \\
n_{e(\gamma) \to \gamma} (E_0, E,t)  & \simeq   &
\left [ n_{e(\gamma) \to e} (E_0, E,t) \right ]_A \times
g_1 \left [\overline{s} \left (\frac{\varepsilon}{E_0}, t \right )
, \frac{E}{\varepsilon}\right ]
\end{eqnarray}
with $\overline{s}(x,t)$   given by  (\ref{eq:sa}).
These   expressions 
combine the solution of the   the  shower equations
in approximation~A  with the functions  $p_1(s,x)$ and 
$g_1(s,x)$  introduced   in section 2.2   as part of the
``elementary solutions''  to the shower equations.
For $E \gg \varepsilon$   the solution 
coincides with the one  obtained   in approximation~A, while 
for  $E  \lesssim \varepsilon$
the spectra have  approximately the same shape  
of the ``elementary solution''  that corresponds to
$\overline{s}(\varepsilon/E_0, t)$.

In approximation~B,  it in possible and  natural 
to consider the value 
\begin{equation}
 s =  
 \overline{s}\left ( \frac{\varepsilon}{E_0}, t \right )  =  
\frac{3 \, t }{t + 2 \, \ln (E_0/\varepsilon)}
\label{eq:ss}
\end{equation}
as  {\em the} age of the shower. 

The total electron size of the shower $N_e(E_0, t)$ obtained
integrating over all energies  is  well approximated  by  the
expression:
\begin{equation}
N_{\gamma (e)  \to e} (E_0,t) =
\frac{1} {\sqrt{2 \pi}}
\left [
\left (\frac{E_0} {\varepsilon} \right )^s
~\frac{K_1 (s, -s)}{s}
~ \frac{G_{\gamma (e) \to e}(s) } 
{\sqrt{\lambda_1^{\prime \prime}(s) \, t} }
~e^{\lambda_1 (s) \, t}  
\right ]_{s = \overline{s}(\varepsilon/E_0, t)} 
\label{eq:nint1}
\end{equation}
where we have   used the fact that  integration over energy
is dominated by $E \sim \varepsilon$ and    the result
(\ref{eq:pint}).
It can be easily  seen, that the  maximum of  the size
coincides with the condition $\lambda_1(s) = 0$,
that implies  $s=1$ and, 
solving equation (\ref{eq:ss}):
\begin{equation}
 t_{\rm max} \simeq \ln \frac{E_0}{\varepsilon}
\label{eq:tmax0}
\end{equation}

Energy conservation  in approximation~B  
can be  expressed with the equation:
\begin{equation}
 \int_0^{E_0} dE~E ~n_e (E_0,E,t) +
 \int_0^{E_0} dE~E ~n_\gamma (E_0,E,t) 
= E_0 -  \varepsilon ~
 \int_0^t
 dt^\prime ~ N_e (E_0, t^\prime)
\label{eq:en-cons}
\end{equation}
The left--hand  side  of this  equation is the energy contained
in the shower particles at depth $t$, while
the   second  term  in the  right--hand  side  gives the energy
dispersed in the medium   by the electrons  as  ionization.
Equation (\ref{eq:en-cons}) also implies:
\begin{equation}
\varepsilon ~ \int_0^\infty
 dt ~ N_e (E_0, t) = E_0
\label{eq:en-cons1}
\end{equation}

\begin{figure} [hbt]
\centerline{\psfig{figure=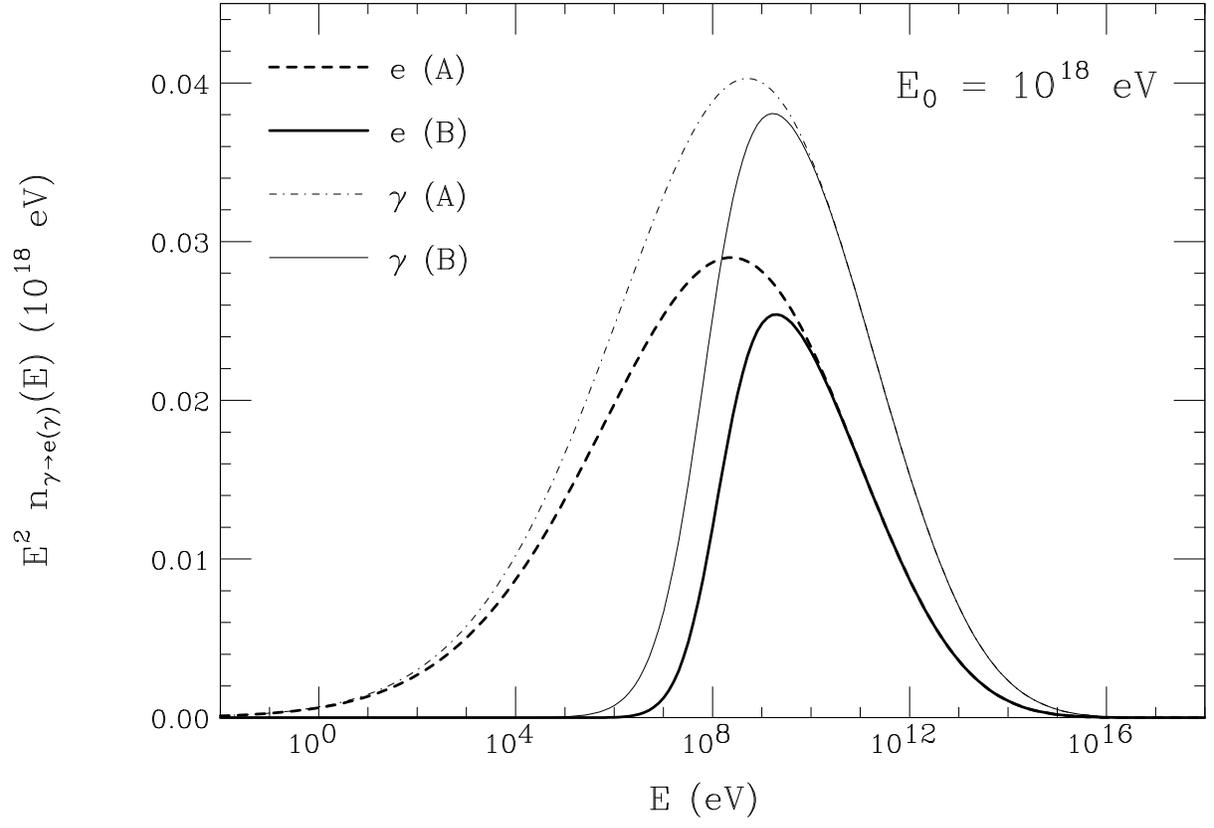,angle=90,width=16.cm}}
\caption {\footnotesize 
Comparison of the 
approximation~A and approximation~B solutions 
for the electron and photon spectra 
in  the shower  generated  by a primary
photon of  energy $10^{18}$~eV
at shower maximum ($t=23.7$).
The area  below each curve 
is  proportional to the amount of  energy   transported by
each particle.
\label{fig:apprb}  }
\end{figure}

An  example of the $e$ and $\gamma$ 
spectra calculated in  approximations A and B
for a photon of initial  energy $10^{18}$~eV
at shower  maximum is  shown  in fig.~\ref{fig:apprb}.
The  two  solutions  coincide
for  $E \gg \varepsilon \simeq 81$~MeV, 
but  deviate from each other  at lower energy.
The spectra  in approximation B  are strongly suppressed
below the critical  energy.
The sum of the areas below  the
curves  are  proportional   to the energy carried by
each particle type.
The curves  of the  approximation~B   solution 
enclose a smaller area  because a part of the  energy 
has been dispersed as  ionization  in the air
(see equations   (\ref{eq:en-cons-A}) and  (\ref{eq:en-cons})).

\begin{figure} [hbt]
\centerline{\psfig{figure=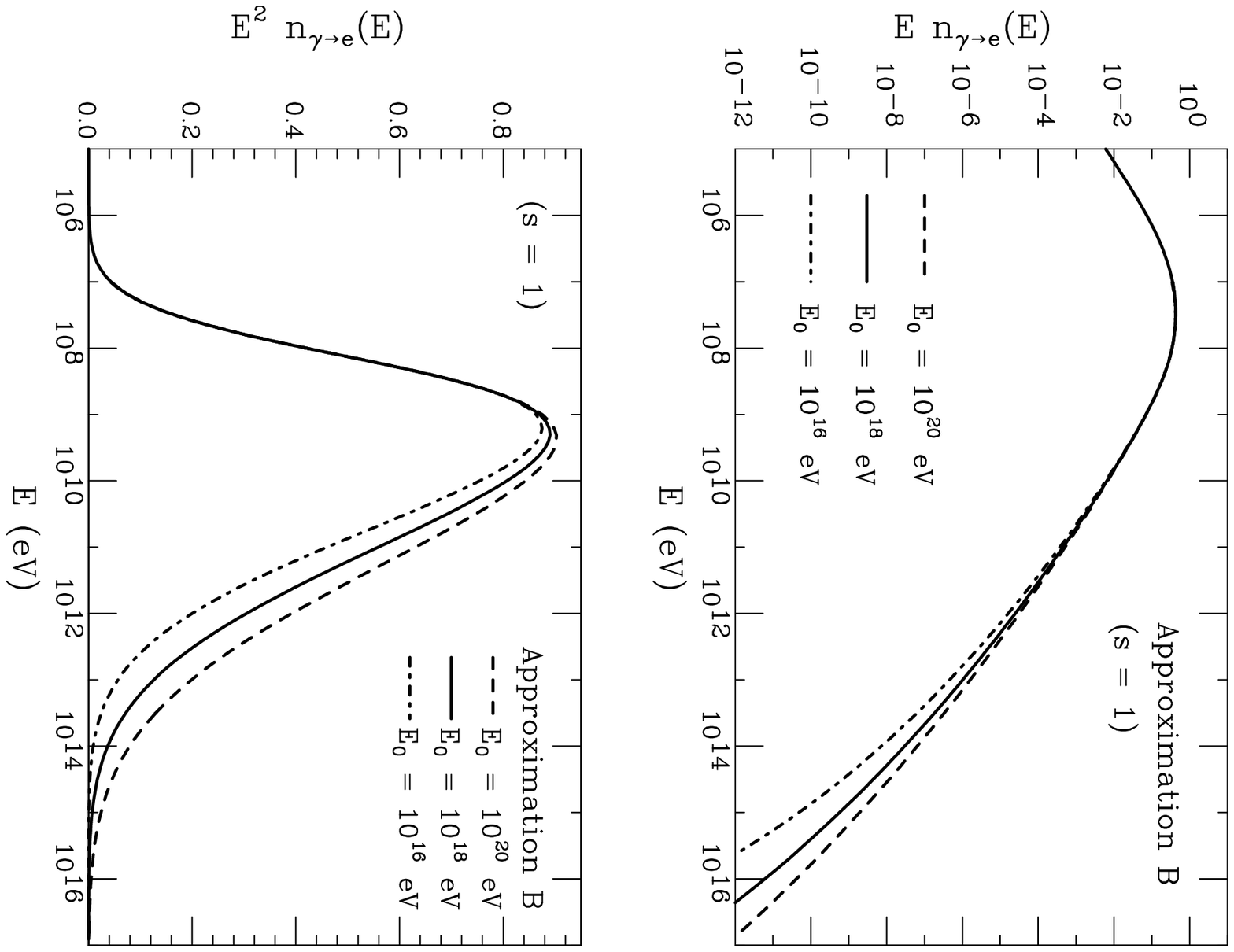,angle=90,width=16.cm}}
\caption {\footnotesize 
Electron  spectra for 
the showers  generated  by  primary
photons of  energy 
$10^{16}$~eV, 
$10^{18}$~eV,  and
$10^{20}$~eV  at  shower maximum
($t=18.6$,
$t=23.7$ and
$t=27.8$).
The normalization is  chosen  so that the spectra
are equal at $E = \varepsilon = 81$~MeV.
The top panel  shows the  spectra
in the form  
$E \; n(E)$ versus $E$;
the bottom panel  shows the same  spectra
in the form 
$E^2 \;  n(E)$ versus $E$.
The similarities and differences  between  the curves illustrates 
the  concept and the limitations of the ``universality''  of the
electron spectra in showers of the same age (in this case $s=1$).
\label{fig:f3}  }
\end{figure}

In  approximation~B  showers generated by different
primaries but having the same    age  $s$  according to the
definition (\ref{eq:ss}) have 
``essentially''  equal  spectra.  
This concept  is illustrated in 
figure~\ref{fig:f3}, that  shows the electron  spectra
at  shower maximum   for   showers  
generated by  photons of
different  energy.
The  spectra are shown in two different
representations. The first  
is of form  ($E\; dn/dE$ versus  $E$),   in this  case
 the area  below the curve is   proportional to
the  electron  multiplicity.  
In the other  representation
the spectra are   shown 
in the form  ($E^2\; dn/dE$ versus  $E$),  in this  case
the area  below the curve is proportional to the  amount of
energy contained in  electrons.
In showers  of the same age   most of the particles
have  coincident spectral  shapes, however the
distributions of  the  highest 
energy particles differ.  High $E$ particles
account for a  significant  fraction of the energy
contained in the shower,  and   are the reason why the evolution
with $t$    of a shower is  not uniquely defined by
the age, but depends also on the shower energy
(or  equivalently of the position $t(s)$ where
the age $s$ is  achieved.

\vspace{0.2 cm}
To summarize the results of  this section: 
the  explicit calculation of the average  development of  purely
electromagnetic  showers  indicates  that   it is possible 
to   define  a   shower age $s$:
\begin{equation}
s \simeq \lambda_1^{-1}  
\left [
\frac{1}{N_e} \frac{dN_e}{dt}
\right ] 
 \simeq \frac{3 \, t}{t + 2 \, \ln(E_0/\varepsilon)}
 \simeq \frac{3 \, t}{t + 2 \, t_{\rm max}}
\label{eq:ssa}
\end{equation}
In  showers of  the same age the 
electrons and photons 
around and below the critical  energy $\varepsilon$
(that  dominate the total  number of particles
in the shower) have the spectra of 
the same shape.
The energy spectra   differ   at larger energy.

\section{The Greisen profile}
\label{sec:greisen}
The average longitudinal development of 
a purely electromagnetic shower 
generated  by  a photon or electron of
energy $E_0$  can be accurately described 
by a  simple  analytic  expression 
introduced  by Greisen \cite{Greisen1}:
\begin{equation}
N_{\rm Greisen}(E_0,t) = \frac{0.31}{\sqrt{\ln (E_0/\varepsilon)}}~
\exp \left [
t \left (
1 - \frac{3}{2} \log \left ( 
\frac{3t}{t + 2 \, \ln (E_0/\varepsilon)}
\right )
\right )
\right ]
\label{eq:greisen}
\end{equation}
The  ``Greisen Profile''
 is   essentially identical
to the  more complex
expression    given in (\ref{eq:nint0}).
The  derivation    \cite{Greisen1} of equation (\ref{eq:greisen}) 
is simple and  instructive, and     requires 
the intelligent ``recombination'' of some   the 
the results obtained above.

The starting point  of the derivation  is the remark
that the    saddle point  solution for  the  total 
electron size (\ref{eq:nint1})   indicates the  
approximate  validity of the relation:
\begin{equation}
\frac{dN_e(t)}{dt}
= \lambda_1(s)\; {N_e(t)} ~.
\label{eq:longdiff}
\end{equation}
with   the age  $s$   given by (\ref{eq:ssa}).
One can now substitute for $\lambda_1(s)$ the 
approximation $\overline{\lambda}_1(s)$ given
in (\ref{eq:lambda1-greisen}) and  rewrite 
equation (\ref{eq:longdiff}) as:
\begin{equation}
\frac{dN_e(t)}{dt} =
\lambda_1 (s) \;  N_e(t)  =
\frac{1}{2} \left [
 \frac{3t}{t + 2 \, t_{\rm max}}  -1 
- 3 \, \log 
\left ( 
\frac{3t}{t + 2 \, t_{\rm max}}
\right )
\right ] ~ N(t) ~. 
\label{eq:greisen1}
\end{equation}
The  solution of this   differential equation 
for  the boundary condition 
$N(t_{\rm max}) = N_{\rm max}$ is readily found as:
\begin{equation}
N_e(t) = N_{\rm max}~e^{-t_{\rm max}} ~
\exp \left [
t \left (
1 - \frac{3}{2} \log \left ( 
\frac{3t}{t + 2 \, t_{\rm max}}
\right )
\right )
\right ]
\label{eq:greisen_integration1}
\end{equation}
The   normalization   is    fixed  observing
that the size at maximum for  an electromagnetic  shower
can be  obtained   inserting the value $s=1$ in
equation  (\ref{eq:nint1}) with the   result:
\begin{equation}
N_e^{\rm max}  (E_0) =
\frac{0.31}
{\sqrt{\ln (E_0/\varepsilon)}}
~\frac{E_0}{\varepsilon}
\end{equation}
where we have used:
\begin{equation}
\frac{1} {\sqrt{2 \pi}} 
~\frac{G_{\gamma \to e} (1)}{\sqrt{\lambda_1^{\prime\prime}(1)}}~ 
K_1 (1, -1) 
= \frac{1} {\sqrt{2 \pi}} 
~\frac{G_{e \to e} (1)}{\sqrt{\lambda_1^{\prime\prime}(1)}}~ 
K_1 (1, -1)  \simeq 0.31
\end{equation}
Substituting this results in (\ref{eq:greisen_integration1})
one  obtains the  final result  (\ref{eq:greisen}).
Expressions (\ref{eq:nint1})  and (\ref{eq:greisen}) ``look''
different from  each other but are essentially  coincident
numerically.

A test  of  the accuracy  of the 
Greisen profile  solution 
can  be  performed  verifying 
 energy conservation   using equation  (\ref{eq:en-cons1}).
This   energy conservation condition is  satisfied to better
than 4.5\% in the   broad energy range from
a few GeV to 10$^{20}$~eV.

In summary,   the Greisen profile  (\ref{eq:greisen})   and 
the    expression for the  age in   (\ref{eq:s0})
(or \ref{eq:ssa}) are equivalent to each other.
The age definition (\ref{eq:s0}) implies
 that the shower  develops
with the Greisen  profile (\ref{eq:greisen1}),
 viceversa the Greisen  profile 
implies the  simple    functional  dependence
for the age of equation (\ref{eq:s0}).
The Greisen  profile and the ``Greisen age''  (\ref{eq:s0})
are (via the  mapping $\lambda_1(s)$)
the integral  and the  derivative of each other:
\begin{equation}
s(t, t_{\rm max}) = \frac{3 \, t}{t + 2 \, t_{\rm max}} 
~~ \Longleftrightarrow ~~ 
N(t)  = _{\rm Greisen} (t,t_{\rm max})
\end{equation}

A recent paper  by Schiel and Ralston \cite{Schiel:2006vf}
has  complained  that  while the Greisen profile
is  shown and discussed  many articles and  textbooks,
a   full derivation is   missing. 
The authors  of  \cite{Schiel:2006vf}  make an 
attempt to ``reverse--engineer'' the steps
performed  orginally performed by Greisen to obtain this  result, 
and arrive at the surprising and erroneous  conclusion  that
the Greisen profile was  ``likely motivated by early numerical 
work in a time predating high--speed computer''.
These  comments miss  the  essential point
that the Greisen  profile 
is the result of the exact integration of  
a  well defined  (albeit product  of some approximations)
differential  equation.

The  work  \cite{Schiel:2006vf}  contains  also a  serious error
when it argues that
performing the  derivative of the Greisen
profile  with respect to the critical  energy 
and setting   the   value   $\varepsilon \to E$, one obtains
the electron  differential spectrum at the energy $E$
for a shower of  primary energy $E_0$ at the depth $t$.
This  error  originates  in a 
confusion  between  approximation~A and approximation~B 
for the shower equations, or perhaps  more accurately
in using  the  crude  assumption 
to  consider  approximation~B as nothing else
than the introduction of a  ``sharp''  cutoff  
for the  electron spectra of approximation~A  
for $E\le  \varepsilon$.

In fact performing the  ``trick''   of  the derivative with respect
to $\varepsilon$   in the 
energy interval  $ E \gg \varepsilon$.
yield  an interesting  result that is  proportional 
(but  not equal) to the electron 
differential spectrum:
\begin{equation}
\left .
-\frac{\partial N_{\rm Greisen} (E_0,\varepsilon, t)}
{\partial \varepsilon}
\right |_{\varepsilon = E} \simeq 
K(s,-s)~n_e (E_0, E, t)
\label{eq:error}
\end{equation}
(with $s=s(t,E_0)$ according to equation (\ref{eq:ssa})).
This  result can  be obtained  comparing equations
(\ref{eq:nint0}) and  (\ref{eq:nint1}).
When $E$   aproaches  $\varepsilon$
the   interpretation of  the  derivative in  (\ref{eq:error})
ceases to be  valid,  and the  electron  spectrum 
takes  the ``universal''  (age dependent)
shape   $\propto  p_1(s,E/\varepsilon) \; E^{-(s+1)}$
with appropriate normalization.

This  last  exercise  is    however instructive
 in the sense  that it illustrates an important point. 
The age  $s$   determines  the   shape of the energy spectra
of the ``bulk''  of the electrons  and photons in the shower,
but the distribution   of the   high energy particles
 depends  on additional parameters.
For the average development of  an electromagnetic  shower
the ony  additional  parameter is the  primary particle
energy  $E_0$  (or equivalently   the position of maximum
$t_{\rm max} \simeq \ln (E_0/\varepsilon))$.
The  high energy particle content 
is crucial  for  the overall  development  of the shower. 
The  Greisen  profile  implies 
that at each level $t$ the shower  contains 
a spectrum of high energy particles 
that is  consistent with  the shape 
of the   development.
This high energy particle content   is  not determined
by the shower age,  but (for  each age) depends  also on the 
primary particle energy $E_0$.
The high energy ($E \gg \varepsilon$)  electron  spectrum 
can be ``extracted''   from the shower  profile
via equation (\ref{eq:error}).

\section{Universality}
As  discussed in the introduction,
the   recent   works 
of  Giller et al. \cite{Giller} and Nerling et al. \cite{Nerling}
have    shown  that  
for  the same   shower age 
(using  the  definition of  equation (\ref{eq:s0})),
{\em individual}  showers
of {\em hadronic} primaries  have  
electron spectra of the same  shape.
This  result is  clearly an  important  generalization of
the   result obtained   in the previous section
that  the {\em average}  development
of purely {\em electromagnetic} showers.

In the following we want to:
\begin{enumerate}
\item   Show that
the  shapes 
of the electron spectra 
 calculated by montecarlo 
for individual hadronic  showers
in \cite{Giller,Nerling}
are essentially identical to the 
Rossi--Greisen  shapes,  calculated
for the  average development of  electromagnetic  showers,
and  given for  each age $s$ by the expression
$p_1(s,E/\varepsilon) \, E^{-(s+1)}$. 

\item  Argue that this ``universality'' 
 result is correct and expected, but that the 
 definition of shower age  should  be modified from
 equation (\ref{eq:s0}) that strictly speaking
is only applicable to the average development  of electromagnetic
shower, to the much more general  form (\ref{eq:age-def}).

\item Show that also the shape of the photon  energy distribution
and  its normalization  relative to the spectrum of
electrons in the shower are ``universal'' and 
 determined by the shower age.
\end{enumerate}

\begin{figure} [hbt]
\centerline{\psfig{figure=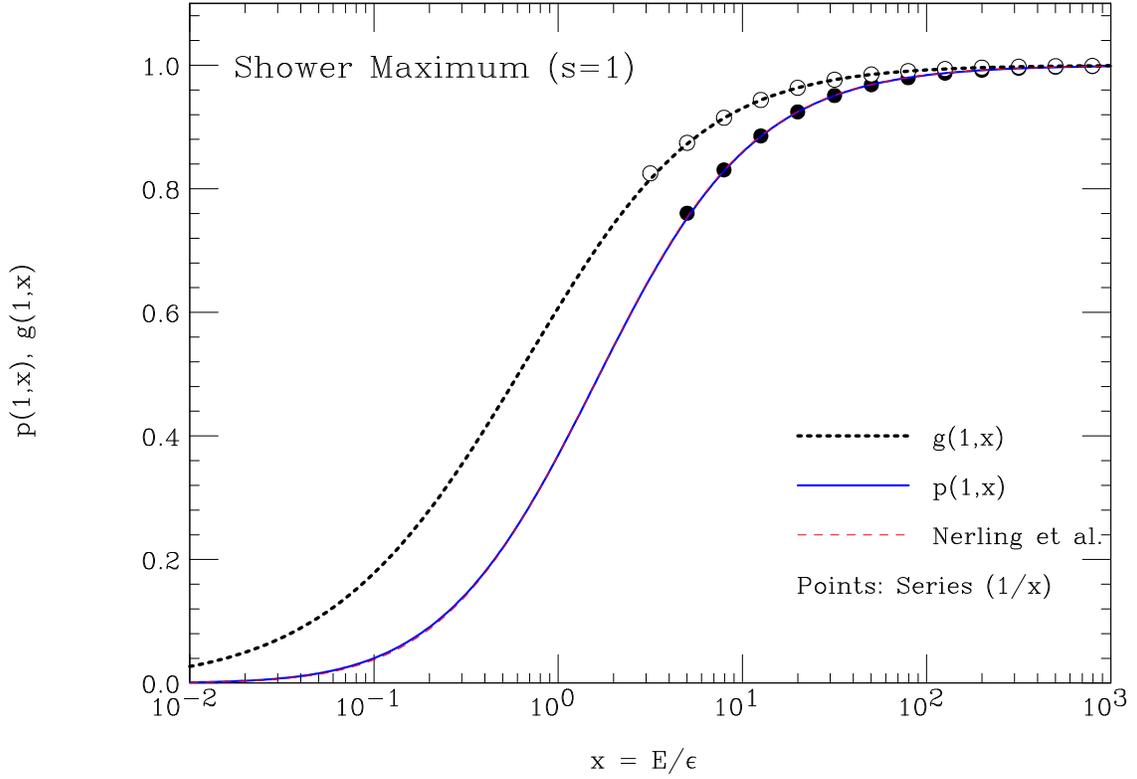,angle=90,width=15.cm}}
\caption {\footnotesize 
Plots of the functions $p(s,x)$ and $g(s,x)$
at  shower maximum ($s=1$). The points  show the results 
obtained  taking the first 4 terms 
the power  series   developments 
(\ref{eq:p-series})  and
(\ref{eq:g-series}).
The dashed line shows the  corresponding  fit of Nerling et al
\protect\cite{Nerling} for the electron spectrum
at  shower maximum.
\label{fig:pp1}  }
\end{figure}

 As an example  of the  numerical  coincidence 
 of the Rossi--Greisen spectra  of  equation 
(\ref{eq:pow1}) with  the shape of the  electron spectra 
 calculated  with montecarlo methods
 for the  same  value of  the  parameter  $s$, 
 in fig.~\ref{fig:pp1} we compare 
 the function  $p_1(s,x)$   for the value  $s=1$
 with  the equivalent quantity
 (that is $n_e (s,E) \; E^{(s+1)}$) 
 from  the fit  to the electron spectra
 at shower maximum   obtained by 
 Nerling et al. \cite{Nerling}.
 The two  functions  are   nearly coincident,
 and agreement of comparable  quality is obtained for
 all  age  values in  the  phenomenologically
 most important range $ 0.7 \lesssim s \lesssim 1.4$
(for more discussion see appendix~\ref{sec:mc}).
 It is also interesting to note 
 that the parametrization  of   \cite{Nerling}
 for the electron  spectrum  has the form:
 \begin{equation}
 n_e (s, E) = \frac{1}{[E + a_1(s)] \; [E + a_2(s)]^s}
 \label{eq:nerling}
 \end{equation}
 that  has   manifestly    the same 
 asymptotic behavior  
 as the Rossi--Greisen shape
 at   both  low  and  high energy:
 $n_e (E) \to $~ constant  for $E \to 0$, and
 $n_e (E) \propto  E^{-(s+1)}$ for $E \gg \varepsilon$.

 The  fact that the  energy spectra 
 calculated   of  individual hadronic  showers 
 coincide   with  remarkable
  accuracy with    spectral shapes  calculated
 for  the  average development 
 of purely electromagnetic   showers  may appear 
 at first sight surprising, but it is  of course not a 
 simple   numerical coincidence, and has in fact a  natural
 explanation.

 The first simple point  is that in all  shower types
(electromagnetic, hadronic and also  neutino  induced)  
the total number of   charged  particles  is  essentially always 
(with the exception of  very early and  very late stages
of development) dominated  by  electrons
with energy around the critical energy
$\varepsilon$.  In hadronic  shower  this   happens because 
 in each hadronic  interaction  a   large fraction of
the energy is transfered  to   photons  via 
the production and decay of $\pi^\circ$ and $\eta$ 
mesons,  these 
photons  then  generate the  electromagnetic  part of the shower
that  accounts for a  growing fraction of the shower energy,
and   for most of the particles in the shower.

The ``universality''    of the spectra 
in different  showers of the same    age,
can then  be   immediately relating  the ``age'' 
with  the  size slope  $\lambda$ according
to equation (\ref{eq:age-def}),
 and  observing that the arguments outlined 
in the previous  sections  can
be  generalized  to conclude that
that  the size slope $\lambda$ must be associated
to a  well defined shape   for the ``bulk'' of the
electrons and photons and to a well defined  relative normalization 
between the two  populations.

For example, at shower  maximum, 
when  the shower size is ``stationary''
($dN/dt \simeq  0$),   the 
photon and electron  spectra  must  have  spectral shapes
and  relative normalization that  insure this stationarity
of the shower size. 
A  stable  solution  for this  problem
has  been found in the previous  section  and is:
 \begin{equation}
 \left \{ 
 \begin{array}{l c l} 
 n_e (E) & \propto  &  
  E^{-2}
 \; p\left (1, \frac{E}{\varepsilon} \right ) 
\\[0.3 mm]
 n_\gamma (E) & \propto  & 
  1.31~ 
\; E^{-2}
\; g\left (1, \frac{E}{\varepsilon} \right ) 
 \end{array}
 \right .
 \label{eq:ppp}
 \end{equation}
To demonstrate  formally that this in  fact  the general
structure of the  electron and photon spectra 
at shower maximum is in fact not trivial,  however
this is  a  very natural   conclusion,  observing how  the 
result (\ref{eq:ppp}) 
emerge  as the  spectra at shower  maximum  for 
the  average devolopment of the shower
generated  by both electrons 
or photons of arbitrary energy.

Similarly, for each  size slope $\lambda$   the
quantity $\lambda_1^{(-1)}(s)$  can  be
be identified   with  the parameters  that labels
the  shape  of the electron  and photon spectra  according
to equation (\ref{eq:pow2}). 
The  argument 
that we  have  outlined  
to relate the    age and the electron and photon
energy  spectra is independent from 
the nature of the primary particle,  from   its 
energy, and from  the value of  the 
depth where the shower is  measured.

The   argument  however does {\em not}  allow to 
estimate  the age   from a closed  form relation
of type $s(t,t_{\rm max})$ such as equation (\ref{eq:s0}).
The  fact  that this is impossible can  be
 easily illustrated   with the example of  the shower  generated
by a neutrino.  In  this case the   $t_{\rm max}$ of the shower
can of course be arbitrary large, and therefore 
for example  the definition (\ref{eq:s0}) returns
for the entire  shower development $\overline{s} \simeq 1$,
that is of course  meaningless, while  it is physically
transparent  that the age concept    mantain its  validity  and
applicability, and the general   definition (\ref{eq:age-def}) has
no difficulty in dealing with   neutrino induced  showers.

One  may think that the neutrino example   described above is
``artificial''  and that  the problem that emerged
can be  ``solved''   for example 
shifting the origin of the depth measurement   to the
 neutrino interaction point.  However this point
is in most cases  unobservable, and this 
shifting  procedure is 
operationally not well defined  (and would open the  problem
of  performing a  similar  shift  of the origin
of the $t$  measurements also for  hadrons and photon primaries).

Another way of  to  see the    problem 
for a closed form expression   of the 
 age of form  $s = s(t,t_{\rm max})$  is that
such a definition  {\em implies}  via
equation (\ref{eq:longdiff})
the entire   longitudinal  profile of the shower.
For example, as discussed  before,
the expression (\ref{eq:s0}) implies
that the shower  develops
with the Greisen  profile (\ref{eq:greisen1}) (and  viceversa).
In general the longitudinal development
of cosmic  rays showers cannot  be   described
accurately with the form  (\ref{eq:greisen1}), and 
this  failure  implies limitations for  the   approximate
definition (\ref{eq:s0}).

\subsection{The ``Gaisser Hillas''  longitudinal profile}
The  observations of 
 longitudinal profile of high energy showers
obtained with the detection of  
fluorescence  light,
 using the technique pioneered by the Fly's Eye detector, 
and currently in use by the HiRes and Auger collaboration, are
commonly  fitted using 
a 4--parameters   expression 
known as the ``Gaisser--Hillas'' profile: 
\cite{Gaisser-Hillas}:
\begin{equation}
N_{\rm GH}(t) = 
N_{\rm max} \;
\left (
\frac{t - t_0}{t_{\rm max} - t_0}
\right )^{\frac{t_{\rm max} - t_0}{\Lambda}} ~
\exp \left [
\frac{t_{\rm max} - t_0}{\Lambda}
\right ]
\label{eq:gh}
\end{equation}
The maximum of this function 
is  at $t = t_{\rm max}$,  
where  the  size  is equal
to $N_{\rm max}$, 
while $t_0$ and  $\Lambda$   modify the shape.
The Gaisser--Hillas  profile  (\ref{eq:gh})    implies  the   shower age:
\begin{equation}
s =  
\lambda_1^{(-1)} \left [
\frac{1}{N_{\rm GH}(t)} \; 
\frac{ dN_{\rm GH}(t)}{dt} \right ] 
=
\lambda_1^{(-1)} \left [
-\frac{1}{\Lambda} \; 
\frac{(t-t_{\rm max})}{(t-t_0)} \right ] 
\end{equation}
where $\lambda_1^{(-1)}(x)$ is the inverse  function.
of $\lambda_1(s)$.
For completeness we note that the inverse
of the function $\overline{\lambda}_1(s) = \lambda$ 
can has the explicit form:
 $\overline{\lambda}_1 ^{(-1)} (\lambda) = -3 \; {\rm ProductLog} \{-1/3 \, 
\exp [(-1 - 2 \, \lambda)/3]\}$.
To    study the     age near shower  maximum, one 
expand  in a power   series  around
the position of the maximum, 
using as expansion parameter the quantity $\delta$:
\begin{equation}
\delta = \frac{(t-t_{\rm max})}{(t_{\rm max}-t_0)}
\end{equation}
The first terms in the series expansion are:
\begin{eqnarray}
s &  = &
 1 + 
\frac{\delta}{\Lambda} 
-
\left (
\frac{4  \,\Lambda  - 3}{4\, \Lambda ^2} 
\right )
 \; \delta^2 
+
\left (
\frac{8 \, \Lambda ^2 -12 \, \Lambda + 5}{8  \, \Lambda ^3}
\right )
 \; \delta^3   +   \ldots
\label{eq:ser-gh}
\end{eqnarray}
For comparison, the age near shower maximum for the 
 Greisen profile (\ref{eq:greisen1}) 
can be written as  a  power expansion
in the quantity  $\delta^\prime$:
\begin{equation}
\delta^\prime = 
\frac{(t-t_{\rm max})}{t_{\rm max}}
\end{equation}
with the result:
\begin{eqnarray}
s &  = &
\frac{3 \, t}{t + 2 \, t_{\rm max} } =
 1 + 
2 \;
\sum_{k=1}^{\infty}  \; (-1)^{k+1}~ 
\frac{1}{3^k} \;
\left ( \delta^\prime \right )^k
\nonumber \\
& ~ & \nonumber \\
& = & 1 + \frac{2}{3} \; 
 \delta^\prime 
-
\frac{2}{9} \; 
\left ( \delta^\prime \right )^2
+
\frac{2}{27} \; 
\left ( \delta^\prime \right )^3
+ \ldots
\label{eq:ser-greisen}
\end{eqnarray}
A comparison of the expansions (\ref{eq:ser-gh})
and (\ref{eq:ser-greisen})  
shows   explicitely  how  precisely
expression (\ref{eq:ser-greisen})  maps   the true  age of
a shower.

\section{Conclusions and Outlook}
The concept of   the shower age 
can be very useful   for the analysis  of high energy
cosmic ray data.  The essence of the idea is  very  simple
and can  be  summarized  in a nutshell
saying that the $t$--slope  $\lambda$ 
and the $E$--slope  $s$    of a shower are
connected to each other  by a one to one mapping.
The $t$--slope (or size slope)  is
the fractional  rate of change of the 
shower size with increasing depth
($\lambda = N^{-1} \, dN/dt$).  The $E$--slope
(or energy slope) is the  integral  slope of  the  
(power law)  energy spectra of
photons or electrons   above the critical energy.
The mapping   between   $\lambda$ and $s$   is given by
$\lambda = \lambda_1(s) \simeq (s -1 - 3 \, \ln s)/2$.
The spectra of photons and electrons have a more
complex  shape  around and below  $\varepsilon$ that is also
determined  by $s$ (or $\lambda$), and have 
a  relative normalization
also determined by $s$ (or  $\lambda$).
It is remarkable that the  electron and photon spectra 
that correspond to different $s$ (or $\lambda$)   have been calculated
accurately  with analytic  methods   by  the pioneers Rossi and 
Greisen many decades  ago. 

These properties of ``universality''   extend
to the  angular and lateral  distributions of electrons and photons.
This  crucially important  subject is not discussed  here
(see appendix~\ref{sec:lateral}  for  some remarks).

The  definition of age  discussed   here  
is independent  from  the  shape of the longitudinal 
development  of a  shower, and is therefore more general
and accurate that the commonly used    definition
$s \simeq  3 \, t/(t + 2 \, t_{\rm max})$ that 
is  correct  only when  the   shower development
is described by the  ``Greisen profile''.
The average shape (and the fluctuations around this  average)
of  the longitudinal development of high energy cosmic  ray showers
 is determined
by the nature of the primary particles and by the properties
of   hadronic  interactions. 
The  observation of  these  shape  is  a very important
subject for  future   experimental studies.

A  definition of age that depends only on the derivative
of the shower  size  can be applied  also to neutrino--induced 
showers, and  more  generally  is expected
to remain valid for the description 
of  all shower  where the size is dominated  
by  electrons.
This includes showers  generated  by exotic primaries,
or the presence of unexpected  physics (or unexpected
fluctuations)  in the development
of  the showers by  primary particles of known nature.
The search for   events  that have  unusual
longitudinal   developments, such  as   multiple maxima
is  an  interesting  direction of   research. 
 It is  likely (and at least the best   possible 
{\em a priori} assumption) 
that the spectra  of the electromagnetic  component
around and below the critical energy 
will, also  in these cases,  be controled   by  the shower  age.

These ideas   can be  useful in the 
analysis  of high energy  cosmic  ray  observations in 
several ways.
As examples: (i) the knowledge of the   variations of the
electron  energy spectrum   during the evolution of
a  shower can be used to  obtain a 
better reconstruction of the longitudinal profile  of the shower 
in observations   that  use  
fluorescence and/or Cherenkov light   detectors
(see \cite{Unger:2008uq} for more discussion);
(ii) the reconstruction of the  shower 
age  from the lateral  distribution  of  its electromagnetic
component can in principle help in the reconstruction of the energy in 
surface  array  measurements; (iii)  
in case  of hybrid measurements   of the showers,
 the   redundant
measurement of the age (from the  size longitudinal   development 
and the lateral distribution of the electromagnetic  component
at the  ground)  can allow to   disentangle 
a muon component, allowing   composition measurements,
or test of hadronic interaction  models
(see   \cite{Engel:2007cm} for more discussion).

It should  finally  be stressed  that
the  ``universality'' in
the electromagnetic  component of high energy showers,
is clearly   an important  analysis  tool, but 
gives  only a partial information about the shower.
Other information is  contained in  the 
shower muon component,   moreover the shower core,
that  is  essentially  undetected in large area
shower  arrays,   in most cases also
contains  a significant amount of energy.
The   energy contained in the core
must be  ``infered'' from the  information 
obtained at large distances from the  shower axis,  introducing
unavoidably some model  dependence in the energy
reconstruction.

\vspace{0.4 cm} 
\noindent {\bf Acknowledgments.}\\
It is a pleasure to  acknowledge fruitful discussions
with Ralph Engel, Maurizio Lusignoli  and Silvia Vernetto.

\newpage
\appendix
\section{Cross sections for  fundamental processes}
\label{sec:appA}
In this  appendix we list  the expressions 
of the differential cross sections for the
brems\-strahlung and pair production, the fundamental
processes that control the development of electromagnetic showers.
It is  convenient to measure  the  column density  $X$
in units of radiation length $X_0$
(with the notation  $t = X/X_0$).
The radiation length in air \cite{PDG}  is
approximately  36.66~(g~cm$^2$)$^{-1}$.

The probability  per unit 
of radiation length that an  electron of
energy $E_e$  emits  a  photon of  energy $E_\gamma = v \; E_e$ 
has the asymptotic form:
\begin{equation}
\varphi(v) =
\frac{1}{v} ~ \left [ 1 - \left( \frac{2}{3} - 
     2\,b \right) \,
   \left( 1 - v \right)  + 
  {\left( 1 - v \right) }^2 \right ]~.
\label{eq:phi}
\end{equation}
For the   pair  production  process 
$\gamma \to e^-e^+$, the energy distribution
of the final state  electron is:
\begin{equation}
\psi(u) ={\left( 1 - u \right) }^2 + 
  \left( \frac{2}{3} - 
     2\,b \right) \,
   \left( 1 - u \right) \,u + 
  u^2
\label{eq:psi}
\end{equation}
with  $u = E_{e^-}/E_\gamma$.
Integrating over all $u$ values  one obtains
the pair production probability per   radiation length:
\begin{equation}
\sigma_0 = \int_0^1 du~\psi (u) = \frac{7}{9} - \frac{b}{3} ~.
\end{equation}
In the previous  equations 
$b$ depends on the atomic  number  of the medium:
\begin{equation}
b  \simeq \frac{1}{18 \, \log(183 \, Z^{-1/3} ) }
\end{equation}
For air one has $b \simeq 0.0135$.

Important momenta of the functions  $\varphi(v)$ and $\psi(u)$ 
are:
\begin{eqnarray}
A(s) & = &  \int_0^1 ~dv ~\varphi(v)~ \left [
1-  (1-v)^s \right ]
\nonumber \\
& = &  \left( \frac{4}{3} + 
     2\,b \right) \;
   \left (
\frac{\Gamma^\prime(1 + s)}{\Gamma(1+s)}
 + \gamma \right )
+\frac{s\,\left( 7 + 5\,s + 
      12\,b\,
       \left( 2 + s \right) 
      \right) }{6\,
    \left( 1 + s \right) \,
    \left( 2 + s \right) }
\label{eq:aa} \\
& ~ & \nonumber  \\
B(s) & = &
 2 \;\int_0^1 ~du~u^s ~\psi(u)
=\frac{2\,\left( 14 + 11\,s + 
      3\,s^2 - 
      6\,b\,\left( 1 + s
         \right)  \right) }
    {3\,\left( 1 + s \right) \,
    \left( 2 + s \right) \,
    \left( 3 + s \right) } 
\label{eq:bb} \\
& ~ & \nonumber \\
C(s) & = &  \int_0^1 ~dv~v^s ~\varphi(v) 
= 
\frac{8 + 7\,s + 3\,s^2 + 
    6\,b\,\left( 2 + s \right) 
    }{3\,s\,\left( 2 + 3\,s + 
      s^2 \right) }
\label{eq:cc}
\end{eqnarray}
In  equation (\ref{eq:aa})  $\Gamma^\prime(z)/\Gamma(z)$ 
is the digamma function  and $\gamma$ is the Euler gamma constant
$\gamma \simeq 0.577216$.

\section{Shower equations in approximation A}
In this section we sketch a derivation
of the solutions of the shower equations  in approximation A
for  the initial condition
of a monochromatic electron or photon.

The   first step is to introduce the 
Mellin  transforms  $M_e(s,t)$ and $M_\gamma (s,t)$  of 
the  electron and photon spectra  $n_e(E,t)$ and $n_\gamma (E,t)$.
In general 
the  Mellin  transform of the function $f(E)$ is defined as:
\begin{equation}
M_f (s) = \int_0^\infty dE ~E^{s} ~f(E)
\label{eq:mellin}
\end{equation}
with   $s$  a  complex parameter.
The Mellin transform converges in a strip
bounded by two straight lines parallel to the imaginary axis
($s_1 < \Re[s] < s_2$).
The  inverse  transformation is:
\begin{equation}
f(E) = \frac{1}{2\pi i} \int_C ds ~ E^{-(s+1)} ~M_f(s) 
\label{eq:mellin-inverse}
\end{equation}
where the integration path $C$ runs parallel to the imaginary axis 
within  the strip of convergence 
of  $M_f(s)$. 

Applying the operator:
$$
\int_0^\infty dE ~E^s
$$
to the  shower equations 
 (\ref{eq:show1}) and(\ref{eq:show2})   
one  obtains  a   system of  two linear differential  equations
for $M_e(s,t)$ 
and $M_\gamma(s,t)$: 
\begin{eqnarray}
\frac{\partial M_e (s,t)}{\partial t}  & = &
- A(s) \; M_e(s,t) 
+ B(s) \; M_\gamma(s,t) 
\label{eq:mel1}
\\
& ~& \nonumber \\
\frac{\partial M_\gamma (s,t)}{\partial t}  & = &
+C(s) \; M_e(s,t) 
-\sigma_0 \; M_\gamma(s,t) 
\label{eq:mel2}
\end{eqnarray}
The general  solution of this   system can   be easily  obtained
as a linear combinations of the exponential functions
$e^{\lambda_1(s) \, t}$ and
$e^{\lambda_2(s) \, t}$.

The shower  generated  by an electron of energy
$E_0$  at  depth $t=0$ corresponds to 
the initial  condition:
\begin{equation}
\left \{
\begin{array}{l l l}
n_e (E,0) & = &  \delta [E - E_0] \\[0.15 mm]
n_\gamma (E, 0) & = & 0 
\end{array}
\right .
\end{equation}
or:
\begin{equation}
\left \{
\begin{array}{l l l}
M_e (s,0) & = &  \left (E_0 \right )^s  \\[0.15 mm]
M_\gamma (s, 0) & = &  0 
\end{array}
\right .
\label{eq:bound1}
\end{equation}
while the shower  generated  by an initial  photon of energy
$E_0$ corresponds to
the initial  conditions:
\begin{equation}
\left \{
\begin{array}{l l l}
n_e (E,0) & = &  0
\\[0.15 mm]
n_\gamma (E, 0) & = & \delta [E - E_0]  
\end{array}
\right .
\end{equation}
or:
\begin{equation}
\left \{
\begin{array}{l l l}
M_e (s,0) & = &  0  \\[0.15 mm]
M_\gamma (s, 0) & = &  \left (E_0 \right )^s
\end{array}
\right .
\label{eq:bound2}
\end{equation}
Using these  boundary conditions 
one finds the solutions:
\begin{equation}
\left \{
\begin{array}{l l l}
M_{e \to e} (E_0, s, t) & = & \frac{E_0^s}{\lambda_1(s) - \lambda_2(s)}
\; \{
 [\sigma_0 + \lambda_1(s)] \, e^{\lambda_1(s) \, t} -
 [\sigma_0 + \lambda_2(s)] \, e^{\lambda_2(s) \, t}
\}
\\[3 mm]
M_{e \to \gamma} (E_0, s, t) & = & 
\frac{C(s) \; E_0^s}{\lambda_1(s) - \lambda_2(s)} 
\; 
\{
e^{\lambda_1(s) \, t} -
e^{\lambda_2(s) \, t}
\} 
\end{array}
\right .
\label{eq:mellae}
\end{equation}
and
\begin{equation}
\left \{
\begin{array}{l l l}
M_{\gamma \to e} (E_0, s, t) & = & 
-\frac{E_0^s}{C(s)}
\; 
 \frac{[\sigma_0 + \lambda_1(s)] \;
 [\sigma_0 + \lambda_2(s)]}{\lambda_1(s) - \lambda_2(s)}
~ \{
e^{\lambda_1(s) \, t} -
e^{\lambda_2(s) \, t}
\}
\\[3 mm]
M_{\gamma \to \gamma } (E_0, s, t) & = & 
-\frac{E_0^s}{\lambda_1(s) - \lambda_2(s)}
\; \{
 [\sigma_0 + \lambda_2(s)] \, e^{\lambda_1(s) \, t} -
 [\sigma_0 + \lambda_1(s)] \, e^{\lambda_2(s) \, t}
\}
\end{array}
\right .
\label{eq:mellag}
\end{equation}

The  functions  $n_{e,\gamma} (E_0, E, t)$ can be  obtained
inverting the  Mellin  transformation 
using (\ref{eq:mellin-inverse}), or more  explicitely:
\begin{equation}
n_\alpha (E) = \frac{1}{2 \, \pi \, i} ~
\int_C  ds~ E^{-(s+1)} ~M_{\alpha} (s) =
\frac{1}{2 \, \pi \, i} ~
 \int_{s_0 - i \, \infty}^{s_0  + i \, \infty} ds ~ E^{-(s+1)}~M_{\alpha}(s)
\label{eq:mellinva}
\end{equation}
(with the  subscript $\alpha$ than runs over the 4 cases:
($e\to e$), 
($e \to \gamma$),
($\gamma \to e$) and 
($\gamma \to \gamma$)).
This integral  cannot be  done exactly analytically,
however with modern  tools 
it is  trivial to   obtain the  numerical  result
with any desired   level of accuracy.
In fact the  integrand  function 
is  well  defined  (and available in computer libraries)
for all complex values  $s$.
The   imaginary part of the integral  vanishes
while the real  part gives the physically  observable spectra.

Rossi and Greisen  have also shown 
that using  the ``saddle point'' approximation 
it is possible to obtain simple analytic expressions
that are a very good approximation of the  exact  results
for $t$  not too small, and that remain   
very instructive and useful.
The basic idea   behind the saddle
point  approximation is that the
integrand in (\ref{eq:mellinva})
is an analytic  function  in  the variable $s$.  
For any  analytic function
$f(z) = f(x + i \, y)$  one has:
$$
\frac{\partial^2 f}{\partial x^2} + 
\frac{\partial^2 f}{\partial y^2} = 0 
$$
This  implies that if the  function $f(z)$  
has  a minimum  for
a real  value  $\overline{z}$  
when  $z$  runs  along the real  axis,
the  function will then have a maximum at the same point
along a path that is at right angle  with 
respect to the real  axis.
The integral  is then  dominated
by the value of the function  near the maximum,
and  can be performed analytically
approximating the integrand   
with a Gaussian  function, and using the well known
result that if $Q(x)$ is a quadratic form:
$$
Q(x) = q_0 + q_1 \, x  + \frac{1}{2} q_2 \, x^2
$$
with  coefficient $q_2 > 0$,  one has:
\begin{equation}
\int_{-\infty}^{+ \infty} dx~ e^{-Q(x)}
 = 
\sqrt{\frac {2 \,\pi}{q_2}}
~\exp \left [ 
-Q(\overline{x}) \right ]
\label{eq:gaussian}
\end{equation}
(where $\overline{x} = -q_1/q_2$ is the point where
the quadratic form $Q(x)$ is  minimum, and the integrand
is therefore  maximum).

One can  now  apply this  idea to the integral (\ref{eq:mellinva}).
For $t$  not too small, one  can  neglect
the term proportional to $\exp[\lambda_2 (s) \, t]$
in the expressions for the  Mellin   transforms  given
in equations (\ref{eq:mellae})  and (\ref{eq:mellag})
 and rewrite the integrand of  the inverse  transform  as:
\begin{equation}
\frac{1}{2 \pi  i} \;
 E^{-(s+1)} ~M_{\alpha} (E_0, s) =  
\frac{1}{2  \pi  i} \;
\frac{1}{E} \;
G_{\alpha} (s)  ~
\left [
 \left ( \frac{E}{E_0} \right )^{-s} ~ e^{\lambda_1 (s) \, t}
\right ] ~.
\label{eq:mella}
\end{equation}
where the    functions  $G_\alpha (s)$ are:
\begin{eqnarray}
G_{e \to e} (s)  & = & \frac{ [\sigma_0 + \lambda_1(s)] }
                      {\lambda_1(s) - \lambda_2(s)}
\label{eq:g1}
\\
~ & ~ & \nonumber \\
G_{e \to \gamma} (s)  & = & 
\frac{C(s)}{\lambda_1(s) - \lambda_2(s)} 
\label{eq:g2}
\\
~ & ~ &\nonumber \\
G_{\gamma  \to e} (s) & = &
-\frac{1}{C(s)}
\;
 \frac{[\sigma_0 + \lambda_1(s)] \;
 [\sigma_0 + \lambda_2(s)]}{\lambda_1(s) - \lambda_2(s)}
\label{eq:g3}
\\
~ & ~ & \nonumber \\
G_{\gamma \to \gamma } (s) & = & 
-\frac{[\sigma_0 + \lambda_2(s)]}
       {\lambda_1(s) - \lambda_2(s)}
\label{eq:g4}
\end{eqnarray}
In equation (\ref{eq:mella})  the integrand of the inverse Mellin  
transform has been  written
as the product of a  function that  changes  rapidly with $s$
(in square parenthesis)  and a  function  that is  considered
as  slowly varying.
The  part of the function  that is
rapidly varying with $s$ 
has a minimum along the real axis 
for the value  $s$  determined by the  implicit equation:
\begin{equation}
\frac{d}{ds} \left [
\left (\frac{E}{E_0} \right )^{-s} ~ e^{\lambda_1 (s) \, t}
\right ] = 0
\end{equation}
or equivalently:
\begin{equation}
\lambda^\prime (s) \, t  + \ln \left ( \frac{E_0}{E} \right )
= 0~.
\end{equation}
This  equation has an explicit  solution if one substitutes
for   $\lambda_1(s)$ 
the Greisen  analytic  approximation
$\overline{\lambda}_1(s)$   given in (\ref{eq:lambda1-greisen}).
The solution is:
\begin{equation}
s \simeq \overline{s}  
\left ( \frac{E}{E_0}, t \right ) = \frac{ 3 \, t}{t - 2 \, \ln (E/E_0)}
\end{equation}
One can now complete the calculation 
at the saddle point $s = \overline{s}$
approximating the integrand as a gaussian function
and using equation (\ref{eq:gaussian})
with 
$q_2 \simeq   \lambda^{\prime \prime}(s) \, t$
with the  result:
\begin{equation}
n_{\alpha} (E_0, E, t) \simeq
\frac{1}{E_0} ~ \frac{1}{\sqrt{2 \pi}}~\left [ 
\frac {G_\alpha (s)}{\sqrt{ \lambda_1^{\prime \prime}(s) \, t}}
~ \left ( \frac{E}{E_0}  \right )^{-(s+1)} ~
e^{\lambda_1 (s) \, t}
\right ]_{s = \overline{s} (E/E_0, t) }
\label{eq:ndiff}
\end{equation}
The integral  distributions
can  be calculated 
noting  that for $t$ not too small the
integration is  dominated  by $E$ close
to $E_{\rm min}$. Neglecting
the slow  variation of $s$ with $E$ one finds:
\begin{equation}
N_{\alpha} (E_{\rm min}, E_0, t) 
   \simeq 
\frac{1}{\sqrt{2 \pi}}~\left [ \frac{1}{s} \;
\frac {G_{\alpha} (s)}{\sqrt{ \lambda_1^{\prime \prime}(s) \, t}}
~ \left ( \frac{E_{\rm min}}{E_0}  \right )^{-s} ~
e^{\lambda_1 (s) \, t}
\right ]_{s = \overline{s}(E_{\rm min}/E_0, t) }
\label{eq:nint}
\end{equation}

An alternative  method to obtain    equation
(\ref{eq:nint}) is to   observe that if
$F(E_{\rm min})$ is the   integral of 
 the function $f(E)$    for $E > E_{\rm min}$, then
the  Mellin transform  $M_F(s)$ is given by:
\begin{equation}
M_F (s) = \frac{1}{s+1} \; M_f(s+1)
\end{equation}
and   performing the inversion with 
the the saddle point approximation.

\vspace{0.3cm}
For completeness we  note  that  Rossi and Greisen
suggest some  slightly more     complex  forms
for the saddle point solutions  in place of 
equations (\ref{eq:ndiff}) and (\ref{eq:nint})
for a better  approximation with the  exact  solution.
The idea  is to  make a better choice
for   the ``fast varying'' part of the Mellin transform.
This  can be done  introducing the quantity
$m_\alpha$  and    rewriting the decomposition 
(\ref{eq:mella}) as:
\begin{equation}
\frac{1}{2 \pi  i} \;
 E^{-(s+1)} ~M_{\alpha} (E_0, s) =  
\frac{1}{2  \pi  i} \;
\frac{1}{E} \;
G_{\alpha} (s) \, s^{-m_\alpha}  ~
\left [
s^{m_\alpha} ~
 \left ( \frac{E}{E_0} \right )^{-s} ~ e^{\lambda_1 (s) \, t}
\right ] ~.
\label{eq:mella-1}
\end{equation}
The esponent $m_\alpha$   is chosen 
``ad--hoc'' for better quantitative  results.
Rossi  and Greisen \cite{Rossi-Greisen} suggest the values:
\begin{eqnarray}
m_{e \to e} &  = &  m_{\gamma \to \gamma} = 0 \\
m_{e \to \gamma } &  = &  - m_{\gamma \to e} = -\frac{1}{2}
\end{eqnarray}
The  part of the function  that is
considered as  varying rapidly with $s$  is 
indicated in square parenthesis in
(\ref{eq:mella-1})   and 
has a minimum along the real axis 
for the value  $s$  determined by the  implicit equation:
\begin{equation}
\lambda^\prime (s) \, t  + \ln \left ( \frac{E_0}{E} \right )
+ \frac{m_\alpha}{s} = 0~,
\end{equation}
that  can be solved  explicitely if one substitutes
$\lambda_1(s)$  with the Greisen  analytic  approximation.
The solution is:
\begin{equation}
s = \tilde{s}_\alpha  
\left ( \frac{E}{E_0}, t \right ) 
= \frac{ 3 t - 2 \, m_\alpha}{t - 2 \, \ln(E/E_0)}
\end{equation}
One can proceed  with the saddle point solution,  noting
that for the  Gaussian approximation
of the rapidly varying function the parameter
$q_2$ is  now  given by:
$q_2 \simeq  \lambda^{\prime \prime}(s) \, t -  m_\alpha/s^2$.
The differential spectra  can  then be written as:
\begin{equation}
n_{\alpha} (E_0, E, t) =
\frac{1}{E_0} ~ \frac{1}{\sqrt{2 \pi}}~\left [ 
\frac {G_\alpha (s)}{\sqrt{ \lambda^{\prime \prime}(s) \, t -  m_\alpha/s^2}}
~ \left ( \frac{E}{E_0}  \right )^{-(s+1)} ~
e^{\lambda_1 (s) \, t}
\right ]_{s = \tilde{s}_\alpha (E/E_0, t) }
\label{eq:ndiff-1}
\end{equation}
Similarly a more complex expression can be written
to improve on   equation  (\ref{eq:nint}) for the integral spectra. 
For large $E_0$ and large $t$
one can  neglect the  terms  $m_\alpha/s^2$ 
and the    more complex  expressions   for the
differential and integral  spectra
coincide with   the simpler  results
(\ref{eq:ndiff}) and (\ref{eq:nint}). 

\section{Elementary solutions in approximation B}
The form of the solution for the shower  equation
in approximation~B is given in
equation  (\ref{eq:pow1}).
Inserting this expression 
in the shower equations
 one obtains  two  integro-differential  equations
 for the functions $p(s,x)$ and $g(s,x)$:
 \begin{eqnarray}
 \lambda(s) \; p(s, x) & = &
 \frac{2 \, C(s)}{\sigma_0 + \lambda(s)} \;
 \int_0^1 \; du \; u^s \psi(u) ~g \left(s, \frac{x}{u} \right )
 \nonumber \\
 & - & 
 \int_0^1 \; dv \; \varphi(v) \left [p(s,x) - (1-v)^s \, 
 p \left (s, \frac{x}{1-v} \right ) \right ]
 \label{eq:pequa0}
 \\
 & - & (s+1) \frac{p(s,x)}{x} + \frac{\partial p (s,x)}{\partial x}
 \nonumber 
 \end{eqnarray}
 and 
 \begin{equation}
 g(s,x) = \frac{1}{C(s)} ~\int_0^1 dv \; v^s \; p \left (s, \frac{x}{v} 
 \right ).
 \label{eq:gequa}
 \end{equation}
 It is possible to eliminate  the function $g$ in (\ref{eq:pequa0}),
 obtaining an equation for $p(s,x)$:
 \begin{eqnarray}
 \lambda(s) \; p(s, x) & = &
 \frac{2}{\sigma_0 + \lambda(s)} \;
 \int_0^1 \; du \; u^s \psi(u) ~ \int_0^1 \; dv \; v^s \; \varphi(v) ~
 p \left ( s, \frac{x}{u \, v} \right ) 
 \nonumber \\
 & - & 
 \int_0^1 \; dv \; \varphi(v) \left [p(s,x) - (1-v)^s \, 
 p \left (s, \frac{x}{1-v} \right ) \right ]
 \label{eq:pequa}
 \\
 & - & (s+1) \frac{p(s,x)}{x} + \frac{\partial p (s,x)}{\partial x}
 \nonumber 
 \end{eqnarray}

 The  two solutions
 for  $p(s,x)$ are obtained  substituting
 in (\ref{eq:pequa})  the two solutions
 for $\lambda(s)$ given in 
 (\ref{eq:lambda}).
 Note that equation (\ref{eq:gequa}) is a simple integral, 
therefore  from the knowledge of the electron  energy spectrum,
that is the functions $p_{1,2}(s,x)$, it is  trivial
 to obtain numerically the corresponding  photon spectrum,
 while it is not entirely trivial to  solve  numerically
 equation  (\ref{eq:pequa}) for $p(s,x)$.

\subsection{Power Expansion  for the functions  $p(s,x)$ and $g(s,x)$}
\label{sec:expansion}
 A useful result  obtained   by Rossi and Greisen
 is  the  demonstration that the  functions
 $p(s,x)$ and $g(s,x)$ can be expressed as a 
  power  series in $1/x$:
 \begin{eqnarray}
 p(s,x) & = & 
 \sum_j c_j(s) \; x^{-j} 
 =  1 +  \frac{c_1(s)}{x} + \frac{c_2(s)}{x^2} +  \ldots 
 \label{eq:p-series}
 \\
 & ~ & \nonumber  \\
 g(s,x) & = &  
 \sum_j d_j(s) \; x^{-j}
 =  1 +  \frac{d_1(s)}{x} + \frac{d_2(s)}{x^2} +  \ldots
 \label{eq:g-series}
 \end{eqnarray}
 with easily calculable coefficients.
 In fact, inserting these  power series forms  in equations
 (\ref{eq:pequa})
 and
 (\ref{eq:gequa})  one obtains 
 simple  recursive  relations for the coefficients:
 \begin{equation}
 c_n = - \frac{c_{n-1}\; (s+n)}{F(s,n)},
 ~~~~~~c_0 = 1.
\label{eq:cn}
 \end{equation}
 \begin{equation}
 d_n = c_n  ~\frac{C(s+n)}{C(s)}
 \end{equation}
 where the function $F(s,n)$ is: 
 \begin{equation}
 F(s,n) = \lambda(s) + A(s+n) - 
 \frac{B(s+n) \; C(s+n)}{\sigma_0 + \lambda(s)}
 \end{equation}
 and the functions  $A(s)$,  $B(s)$ and $C(s)$ are listed in  appendix~\ref{sec:appA}.
 Clearly a  power  series  in  $1/x$   solution cannot  be used
 for $x$  close or   below unity  (that is for $E$  close or below  the critical energy 
$\varepsilon$),  however this
 expansion     allows to  test
 in its  range of validity solutions  obtained
 with different methods.  

\section{Analytic and Montecarlo solutions}
\label{sec:mc}
The   calculation of the spectra of  electrons
and photons  in  shower of the same age  obtained 
by  Rossi and Greisen   in \cite{Rossi-Greisen}  and
discussed in this  paper  has the limitations  that 
are  intrinsic to   the realistic but simplified 
theoretical framework (approximation~B) that has been used.
This    framework  introduces  several simplifications:
the  cross sections for bremsstrahlung and pair production
have always  the  asymptotic  form
that is  strictly speaking only valid at very high energy,
the electron collision losses  are treated as a simple  
energy independent constant, and Compton  scattering
is  entirely neglected. 
 Also in approximation~B the electron mass
is  neglected  and the  electron spectrum  extends down to
zero  energy.
A montecarlo calculation of the spectral
shapes  can of course avoid all these  limitations
and   is  in principle more accurate,   even if it 
has its own  limitations and difficulties

A  comparison of  the Rossi--Greisen shapes with the 
numerical  results of  \cite{Giller,Nerling},  
shows remarkable   agreement  but  also some small differences,
that could be interesting to explore in more detail.
As an illustration, 
normalizing the  high energy spectra to
$n_e(E,s) \to  E^{-(s+1)}$,   the quantity 
 \begin{equation}
 s \; \int_0^\infty dE ~n_e(s,E) 
 \label{eq:inta}
 \end{equation}
is given by the function  $K_1(s,-s)$   for the 
Rossi--Greisen  calculation.  The numerical  integration
of the Nerling et al  parametrization  \cite{Nerling}
gives   results that differ by   5--10\%
(at shower  maximum the difference is 4.5\%).
 A  comparison of the results  is  shown 
in fig.~\ref{fig:ks}.

\begin{figure} [hbt]
\centerline{\psfig{figure=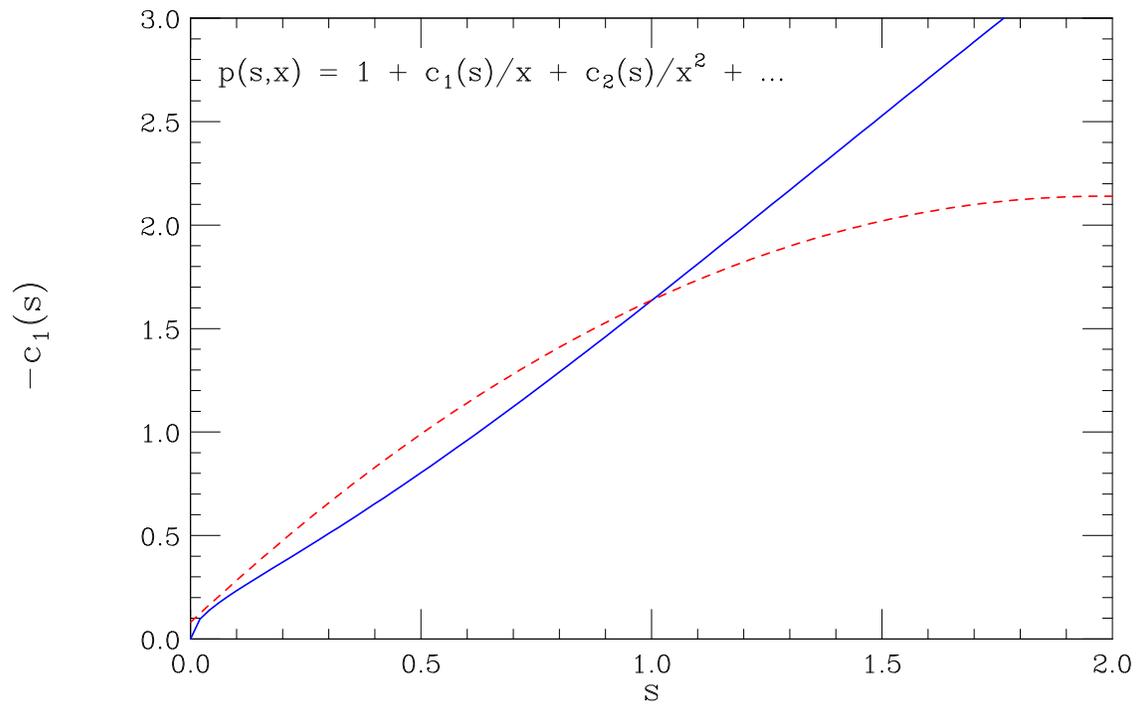,angle=90,width=15cm}}
\caption {\footnotesize 
Plot of the coefficient $c_1(s)$ of
the    expansion  (\protect\ref{eq:p-series})
for the $p_1(s,x)$ function.
The red dashed curve is the coefficient for the
numerical calculation of Nerling et al. \cite{Nerling}.
\label{fig:cc1}  }
\end{figure}

 As  an additional   test we have    calculated 
 the first coefficient in the development:
 \begin{equation}
p_1 \left(s, \frac{E}{\varepsilon} \right ) 
=   E^{s+1} \, n_e(s,E) = 1 + c_1  (s)\, \frac {\varepsilon}{E} + 
 c_2(s) \, \left (\frac{\varepsilon}{E}\right )^2 + \ldots
 \end{equation}
 For the  Nerling parametrization (\ref{eq:nerling})
the first coefficient is $c_1(s) = - [a_1 (s) + s \, a_2(s)]$,
while  for the Rossi--Greisen  solution  the coefficient is
given by (\ref{eq:cn}).
 A comparison of the two  estimate  is  shown in fig.~\ref{fig:cc1}.

The origin of the (small)  differences 
between the analytic and montecarlo solutions   merits
further studies.  A possible explanation
is a  more precise  description of the
physics of the  electromagnetic  interactions in the montecarlo
calculation.

\section{Lateral Distribution}
\label{sec:lateral}
It is intuitive  that in  showers of the same  age  ``most''
electrons  and photons   have not only 
the same  energy  distributions but also 
the same angular  distributions  and,
for the same  density profile of  the medium  where the
showers  are propagating,  also essentially 
equal   lateral distributions around the shower axis.  

The problem of calculating the
electron lateral  distribution  has attracted  considerable
in the past. 
Nishimura  and Kamata  \cite{Nishimura-Kamata} 
solved   numerically   the 3--dimensional shower 
equations in approximation B
to obtain the (energy integrated)
lateral distribution of  electrons  
propagating in a medium of constant  density.
Their result were  fitted by Greisen  \cite{Greisen1} with
the approximate form:
\begin{equation}
\rho_e(s,r) =  N_e \; K(s) \; \frac{1}{r_0^2} 
\; 
\left ( 
\frac{r}{r_0} 
\right )^{s-2} 
\; 
\left ( 1 + 
\frac{r}{r_0} 
\right )^{s- 4.5} 
\label{eq:NKG}
\end{equation}
with  $r_0$ is the Moliere radius
and 
$K(s) = (2 \pi)^{-1} \;\Gamma [4.5-s]/(\Gamma [4.5-2 \, s] \;
\Gamma[s])$ is  
a normalization factor.
After these works  several other authors have   given different
parametrizations of the lateral distribution as
a  function of an age parameter (see
for example  Hillas in \cite{Hillas:1982vn}.

In the view of this author  it is in fact
not possible to have  a  single parametrization 
for  lateral  distribution,  because  it is essential to
consider the   properties of the detector
that  measure the shower at the ground. 
For example the results of Nishimura and Kamata 
refer to the total  number of  electrons  integrated down
to zero energy, however in  most cases
the  detectors of   shower arrays do not sample the
electron number  but an energy deposition, 
and in any case  it is always needed 
to take into account some  contribution from
photons in the shower.   Therefore one needs to combine
appropriately the electron and photon contributions
with the detector response.
An  additional  complications  is of course that
for hadronic  primaries one has  to disentangle the
the electromagnetic and muon components of the shower.

\newpage

\end{document}